\documentclass[11pt,letterpaper]{article}
\usepackage[margin=1in]{geometry}
\usepackage[T1]{fontenc}
\usepackage{lmodern}
\usepackage{amsmath,amssymb}
\usepackage{graphicx,placeins}
\usepackage{booktabs,array,longtable,adjustbox}
\usepackage[font=small,labelfont=bf]{caption}
\usepackage{tocloft}
\usepackage[numbers,sort&compress]{natbib}
\usepackage[hyphens]{url}
\usepackage[hidelinks]{hyperref}

\ifdefined\pdfminorversion
\fi
\graphicspath{{SI_figures/}}

\title{A unified finite-site framework for scaffold--client condensation}
\author{Hamza Coban\textsuperscript{1} and Alfredo Alexander-Katz\textsuperscript{2}\\[0.65em]
  \small\textsuperscript{1}Department of Physics, Massachusetts Institute of Technology, Cambridge, MA, USA\\
  \small\textsuperscript{2}Department of Materials Science and Engineering,\\[-0.15em]
  \small Massachusetts Institute of Technology, Cambridge, MA, USA}
\date{}
\hypersetup{
  pdftitle={A unified finite-site framework for scaffold--client condensation},
  pdfauthor={Hamza Coban and Alfredo Alexander-Katz}
}

\begin{document}
\maketitle
\begin{abstract}
Client recruitment into biomolecular condensates does not determine how stronger binding changes the bulk free-energy advantage of demixing. Here, we use a unified finite-site thermodynamic framework to examine how scaffold--client interactions affect both the response of a fixed-composition mixture and the organization of its ternary phase diagram. The thermodynamic response to increasing client affinity is determined by the difference in heterotypic bonding between the complete equilibrium phase-separated sample and the corresponding homogeneous state. A finite region of the model exhibits positive scaffold--client concentration covariance but negative heterotypic bond contrast: stronger binding reduces the demixing advantage even when both species enrich in the dense phase. This reduction also occurs with independent scaffold self-association and client-binding site pools, without direct displacement of scaffold bonds. A negative response does not by itself imply condensate shrinkage or dissolution. Across composition space, the same finite binding constraints support scaffold-rich and heterotypic coexistence regimes with distinct bond populations. For divalent clients, these regimes can occupy disconnected regions of the phase diagram and reconnect upon increasing client affinity, depending on scaffold cohesion. Critical-point continuation identifies the folds and cusp organizing these changes, while client valence restricts the parameter range over which disconnected coexistence occurs. Explicit-particle Monte Carlo simulations reproduce the qualitative reentrant response and provide finite-size evidence for disconnected coexistence. These results provide a bond-resolved characterization of fixed-parent affinity response and a complementary description of coexistence under specified association constraints.
\end{abstract}

\section*{Introduction}

\begin{figure}[!htbp]
    \centering
    \includegraphics[width=\linewidth]{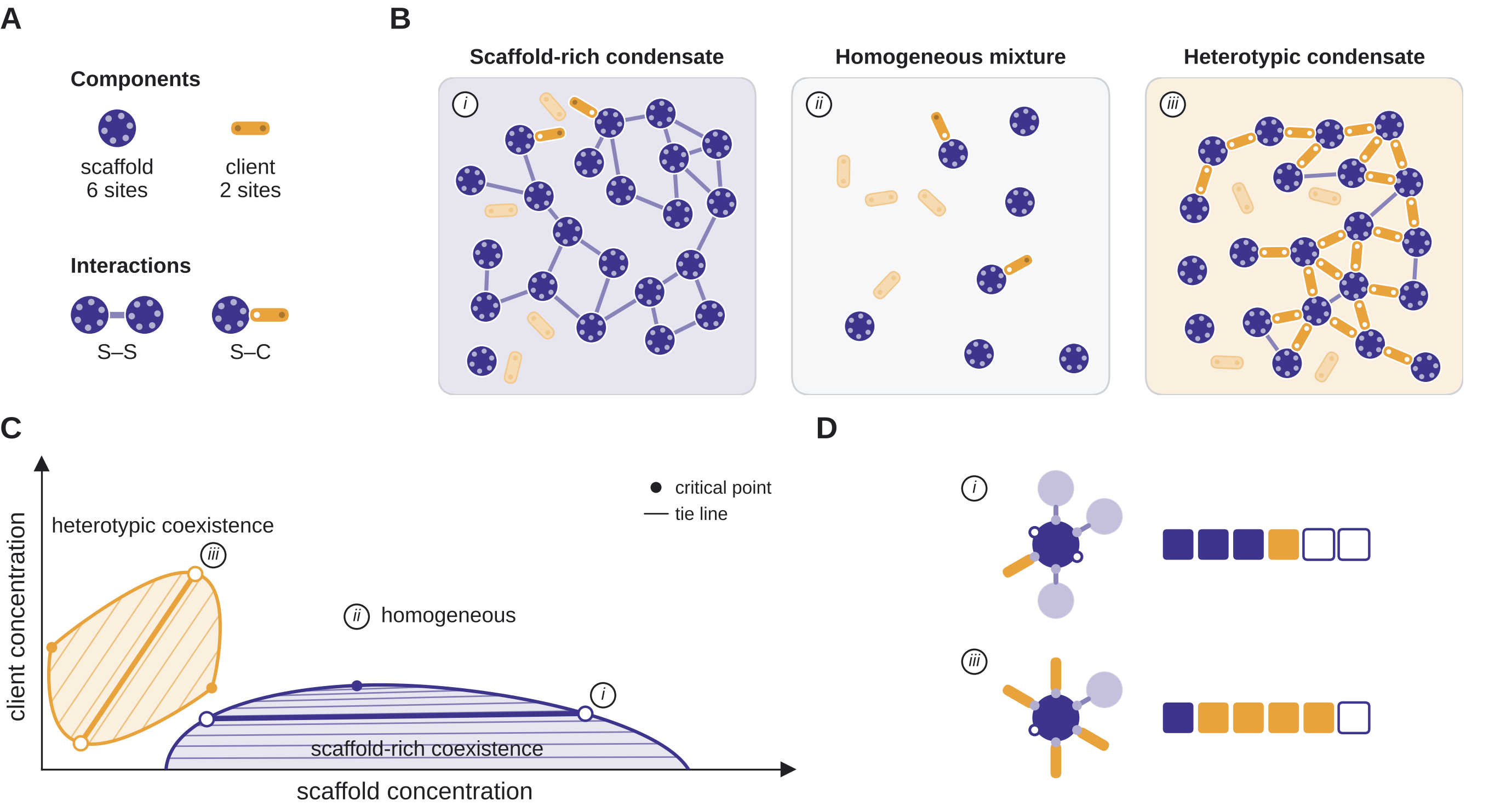}
    \caption{(A) Scaffold--scaffold (S--S) and scaffold--client (S--C) binding, with at most one bond per site. (B) Scaffold-rich, homogeneous, and heterotypic organizations. (C) One possible phase-diagram topology. Tie lines connect coexisting compositions; filled circles mark critical points. (D) Illustrative scaffold-site allocation: S--S bonds (blue), S--C bonds (orange), and unbound sites (unfilled). Unbound sites indicate potential capacity for additional partners outside the present model.}
    \label{fig:intro}
\end{figure}

Biomolecular condensates emerge from collective interactions among proteins, nucleic acids, and other cellular components, with their formation and composition governed by the thermodynamics of multicomponent mixtures \cite{hyman2014,banani2017,choi2020}. Beyond concentrating molecules, condensates can organize interactions that contribute to cellular biochemical processes \cite{lyon2021}. Reconstituted systems have established that multivalent interactions can drive phase separation and that changing interaction valence or affinity alters the conditions for assembly \cite{li2012,banjade2014}. In the minimal scaffold--client network illustrated in Fig.~\ref{fig:intro}A, scaffolds carry several sites that bind either scaffolds or clients, whereas clients bind only scaffolds. Molecular valence specifies the number of available sites; multivalent clients can thus link scaffolds without self-associating. Linker properties and the sequence patterning of interacting residues further influence condensation \cite{harmon2017,wang2018,martin2020}. In multicomponent systems, homotypic and heterotypic interactions jointly determine phase equilibrium, allowing changes in one component to reorganize the composition and interactions of the assembly \cite{riback2020,farag2023,sanders2020}.

The recruitment of additional molecular components can regulate condensation in qualitatively different ways. Nuclear import receptors binding FUS and ubiquitin binding UBQLN2 can suppress phase separation by interfering with interactions responsible for scaffold assembly \cite{yoshizawa2018,dao2018}, while changes in RNA concentration and protein--RNA interactions can produce buffering, dissolution, and reentrant condensation \cite{banerjee2017,maharana2018,alshareedah2019,sanchezburgos2023}. Theory and simulations identify competition for network-forming interactions, client valence, and binding architecture as important determinants of these responses \cite{nguemaha2018,ghosh2019,espinosa2020,joseph2021,ruff2021}. The balance of homotypic and heterotypic binding can favor scaffold-rich condensation, a homogeneous mixture, or heterotypic condensation, while molecular association can occur in all three states (Fig.~\ref{fig:intro}B). For example, NPM1--SURF6 condensates interconvert between homotypic NPM1--NPM1 and heterotypic NPM1--SURF6 interaction networks as protein concentrations and molecular crowding change \cite{ferrolino2018}. These mechanisms motivate a thermodynamic comparison of how a specified mixture responds to a change in affinity.

Thermodynamic linkage provides a complementary description by relating molecular binding to changes in equilibrium \cite{wyman1975,wyman1980}. For condensates, polyphasic linkage describes shifts in scaffold saturation concentration through binding polynomials for ligand binding in the dilute and dense phases \cite{ruff2021linkage}. A distinct comparison arises for a closed mixture in which the total scaffold and client concentrations remain fixed while their mutual affinity is varied. The free-energy advantage of demixing then compares the homogeneous state at the parent composition with the complete phase-separated equilibrium state, including both phases and their volume fractions. Because stronger affinity lowers the free energies of both states, the response depends on their relative use of the regulated interaction. This fixed-parent comparison asks which complete state benefits more from stronger binding.

Experiments using multivalent scaffolds have shown that client recruitment and condensate composition are sensitive to scaffold stoichiometry and client valence \cite{banani2016}. Coexistence compositions and tie-line orientations provide complementary information about molecular redistribution and the interplay of homotypic and heterotypic interactions \cite{deviri2021,farag2023}. A tie line connects the compositions of two coexisting phases; the position of the parent along it determines their relative amounts. Figure~\ref{fig:intro}C sketches one possible topology, with disconnected scaffold-rich and heterotypic coexistence regions separated by homogeneous compositions. Complementary binding capacities provide characteristic composition scales for heterotypic condensation \cite{chattaraj2021,chattaraj2023}, while phase-boundary geometry can quantify solute contributions to multicomponent equilibrium \cite{qian2024,ausserwoger2025}. These descriptions connect measurable concentrations to macroscopic thermodynamics, but the distribution of molecules between phases does not uniquely specify the bonds they form.

Finite-site association thermodynamics retains this molecular information within the free energy used to determine phase equilibrium. Wertheim's thermodynamic perturbation theory provides the association contribution underlying Statistical Associating Fluid Theory (SAFT), developed for associating fluids and subsequently applied to patchy-particle systems \cite{wertheim1,wertheim2,wertheim3,chapman1989}. Molecular valence and binding specificity enter through explicit sites whose occupancies are constrained by partner availability and saturation. Self-consistent mass-action relations determine their allocation among bond types as composition and affinity vary. Sites may support scaffold self-association, bind clients, or remain unbound (Fig.~\ref{fig:intro}D). If the same sites also interact with additional partners, changing this allocation could influence recruitment and biochemical organization, providing a possible connection to condensate function beyond concentration alone. Bond populations and thermodynamic potentials follow from the same free energy, and finite valence and selective association are known to produce unusual critical behavior and coexistence topologies \cite{bianchi2006,delasheras2011}.

Related association-based approaches have also been developed for multivalent biomolecular systems. Associating polymer models show how homotypic and heterotypic sticker interactions, together with solvent compatibility, can produce distinct coexistence regimes and disconnected miscibility gaps \cite{michels2022}. Finite valence and binding stoichiometry can additionally favor finite molecular complexes or suppress macroscopic phase separation despite strong microscopic association \cite{xu2020,zhang2021}. Previous studies have established affinity-dependent reentrance \cite{deviri2021}, shown that ligand enrichment does not by itself classify a regulator as a promoter or suppressor of condensation \cite{ruff2021}, and characterized the interplay of homotypic and heterotypic interactions in multicomponent coexistence \cite{farag2023}. These studies motivate resolving how an affinity perturbation at fixed parent composition changes bond allocation in the complete equilibrium sample.

Here we present a unified finite-site description in which molecular bond allocation, phase equilibrium, and affinity response follow from a common free energy. We derive a criterion for the response to changing scaffold--client affinity at fixed parent composition: the heterotypic bond contrast between the complete separated sample and the homogeneous parent. The framework predicts a finite regime in which scaffolds and clients both enrich in the dense phase, yet stronger heterotypic binding reduces the demixing advantage. Phase-resolved occupancy analysis and an independent-site control identify the mechanism and show that direct site displacement is not required. The same description captures reentrant condensation and distinguishes scaffold-rich and heterotypic coexistence by their bond populations. We determine how client valence and scaffold cohesion control the appearance and reconnection of these coexistence regions.

\section*{Methods}

\subsection*{Finite-site association model}

We consider an incompressible ternary mixture of scaffold, client, and solvent with lattice occupancies $s=\rho_S$, $c=\rho_C$, and $z=1-s-c$, respectively. Each species occupies one lattice cell, so the model represents a coarse-grained mixture without molecular-size asymmetry. Each scaffold carries six equivalent sites that form either scaffold--scaffold ($S$--$S$) or scaffold--client ($S$--$C$) bonds. A $C_n$ client carries $n$ sites that bind only scaffolds. Each site forms at most one bond, and client--client association is absent.

We combine the incompressible lattice mixing free energy with the first-order Wertheim thermodynamic perturbation theory (TPT1) association contribution underlying Statistical Associating Fluid Theory (SAFT) \cite{wertheim1,wertheim2,wertheim3,chapman1989}. The association contribution accounts for finite binding capacity through self-consistent unbound-site fractions, while the lattice reference describes the mixing entropy of scaffold, client, and solvent. The reduced Helmholtz free-energy density is
\begin{equation}
\beta f(s,c)
=
s\ln s+c\ln c+z\ln z
+6s\,g(X_S)+nc\,g(X_C),
\label{eq:free_energy}
\end{equation}
where $g(X)=\ln X-X/2+1/2$. The unbound-site fractions $X_S$ and $X_C$ satisfy the mass-action relations
\begin{align}
X_S^{-1}
&=1+6sX_S\Delta_{SS}
+ncX_C\Delta_{SC}, \quad
X_C^{-1}
=1+6sX_S\Delta_{SC},
\label{eq:mass_action}
\end{align}
with association parameters
$\Delta_{ij}=v[\exp(\beta\epsilon_{ij})-1]$.
Unless otherwise indicated, $v=1/8$ and
$\beta\epsilon_{SS}=1.6$.

\subsection*{Bond allocation and phase equilibrium}

The scaffold--scaffold and scaffold--client bond densities follow from the same mass-action solution:
\begin{align}
b_{SS}
&=\frac{(6sX_S)^2\Delta_{SS}}{2},
\qquad
b_{SC}
=(6sX_S)(ncX_C)\Delta_{SC}.
\label{eq:bond_densities}
\end{align}
The fractions of scaffold sites engaged in $S$--$S$ bonds, $S$--$C$ bonds, or remaining unbound are $2b_{SS}/(6s)$, $b_{SC}/(6s)$, and $X_S$, respectively. We also define the heterotypic fraction of the bond population as $q_{SC}=b_{SC}/(b_{SS}+b_{SC})$. Because the bond types consume different numbers of scaffold sites, this quantity differs from the heterotypic scaffold-site fraction and does not measure network connectivity.

The scaffold and client chemical potentials and osmotic pressure are obtained from the free energy:
\begin{align}
\beta\mu_S
&=\ln(s/z)+6\ln X_S,
\quad
\beta\mu_C=\ln(c/z)+n\ln X_C,
\quad
\beta\Pi=-\ln z-b_{SS}-b_{SC}.
\label{eq:phase_equilibrium}
\end{align}
Thus, the mass-action occupancies determine both the populations of individual bonds and the thermodynamic quantities governing coexistence. Two-phase equilibrium requires equality of both chemical potentials and osmotic pressure. For coexisting phases $\alpha$ and $\gamma$, the parent composition satisfies the lever rule
$(s_0,c_0)=\phi(s_\alpha,c_\alpha)+(1-\phi)(s_\gamma,c_\gamma)$, where $\phi$ is the volume fraction of phase $\alpha$.

We quantify the thermodynamic advantage of demixing at fixed parent composition by
\begin{equation}
G(s_0,c_0)
=
\beta f_{\mathrm{hom}}(s_0,c_0)
-\beta f_{\mathrm{eq}}(s_0,c_0),
\label{eq:G}
\end{equation}
where $f_{\mathrm{eq}}$ is the lower convex envelope of the free-energy surface. Thus, $G>0$ within a coexistence region and $G=0$ for a stable homogeneous parent. The homogeneous reference is the uniform free-energy branch at $(s_0,c_0)$ and may be metastable or locally unstable. This bulk free-energy difference measures the thermodynamic preference for demixing at a specified composition. Fluctuation amplitudes, nucleation barriers, and kinetic rates require additional information; condensate abundance follows separately from the phase compositions and lever rule.

For a tie line, the relative client-to-scaffold transfer is characterized by $t=(c_\gamma-c_\alpha)/(s_\gamma-s_\alpha)$. The complementary heterotypic binding capacities are balanced at $nc=6s$, corresponding to $c/s=6/n$. This ratio provides a reference composition scale for heterotypic association but does not prescribe a coexistence composition or tie-line direction.

Coexistence regions were obtained from the lower convex envelope of the free-energy surface and refined by enforcing equality of chemical potentials and osmotic pressure. Interior critical points were identified by a vanishing Hessian eigenvalue and third directional derivative of the free energy along the corresponding soft mode. Continuation in $(\epsilon_{SS},\epsilon_{SC})$ was used to locate folds of the critical-point locus and their meeting point at a cusp. Coexistence branches were traced independently, and every reported tie line was checked by comparing its common tangent plane with the free-energy surface across composition space. Free-energy scans at fixed client chemical potential spanning both folds tested for competing three-phase states at $\beta\epsilon_{SS}=1.6$.

To examine whether a negative fixed-parent affinity response requires direct competition between $S$--$S$ and $S$--$C$ bonds, we also consider a scaffold with independent binding-site pools. In this architecture, six sites form only $S$--$S$ bonds, while another six form only $S$--$C$ bonds. The separation preserves the pure-scaffold free energy while preventing client binding from directly displacing scaffold self-association. The association free energy becomes
\begin{equation}
\beta f_{\mathrm{assoc}}
=
6s[g(X_A)+g(X_B)]+nc\,g(X_C),
\end{equation}
where $X_A$ and $X_B$ are the unbound fractions of the scaffold self-association and client-binding site classes, respectively. Their mass-action relations are
\begin{align}
X_A^{-1}
&=1+6sX_A\Delta_{SS},
\quad
X_B^{-1}=1+ncX_C\Delta_{SC},
\quad
X_C^{-1}=1+6sX_B\Delta_{SC}.
\label{eq:separate_pool_mass_action}
\end{align}

The two architectures are compared at the same $\Delta_{SC}$ and have identical client-binding capacity on an isolated scaffold. Because the independent-pool scaffold contains more association sites in total, this comparison tests whether direct site displacement is necessary for a negative affinity response. It does not isolate or quantify the contribution of site competition to the response of the shared-site model.

The Supporting Material provides detailed derivations, phase-diagram checks, and Monte Carlo methods and analyses.

\section*{Results}

\subsection*{Increasing client affinity redirects a fixed sample between distinct coexistence states}

\begin{figure}[!htbp]
    \centering
    \includegraphics[
        width=\linewidth,
        height=.68\textheight,
        keepaspectratio
    ]{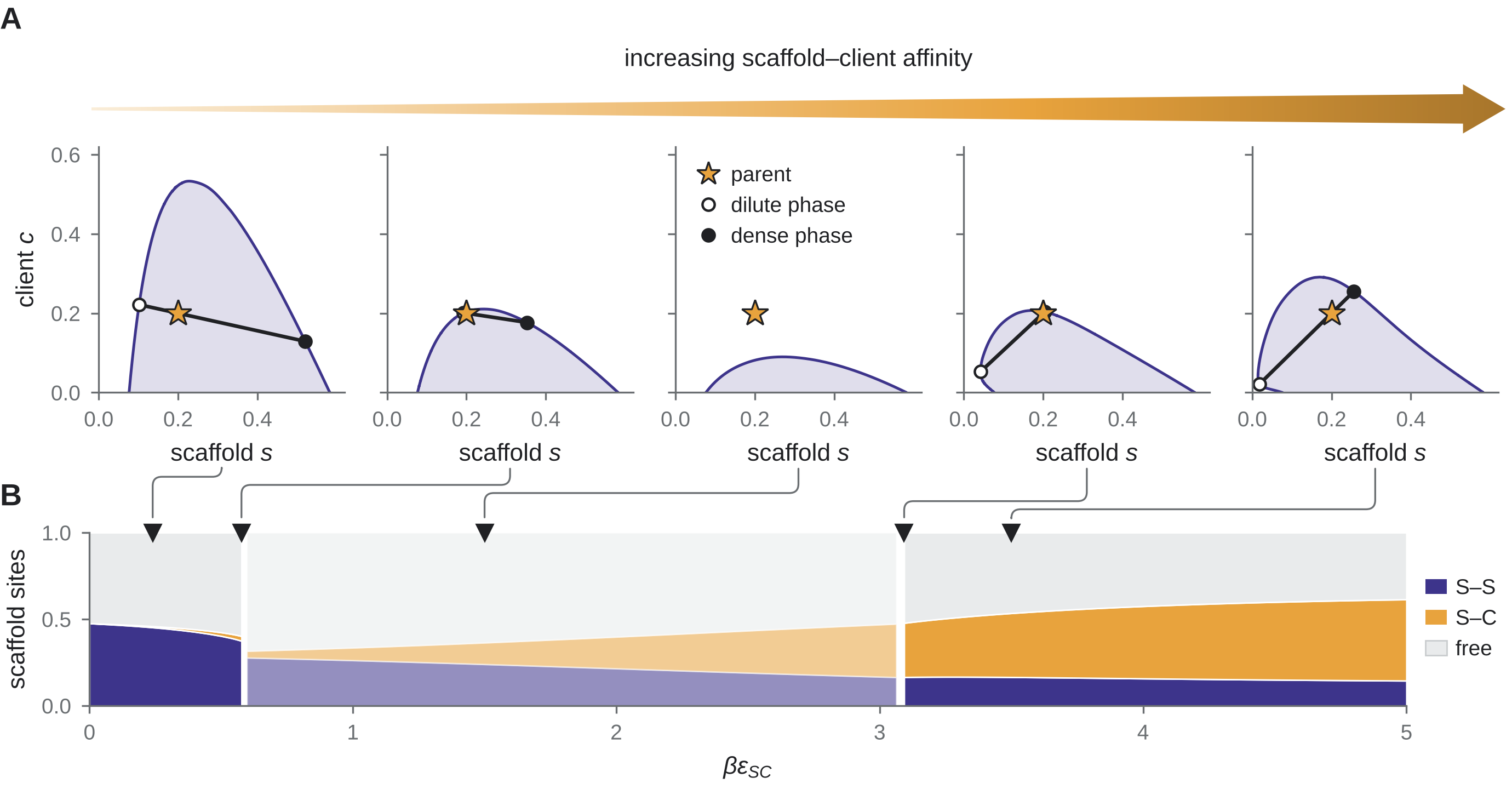}
    \caption{(A) Coexistence regions of the $A_6+C_3$ system at $\beta\epsilon_{SS}=1.6$ and five scaffold--client affinities, $\beta\epsilon_{SC}=0.24$, $0.577$, $1.5$, $3.093$, and $3.5$ (left to right). The orange star marks the fixed parent composition $(s_0,c_0)=(0.20,0.20)$, the shaded regions indicate two-phase coexistence, and black segments denote equilibrium tie lines connecting the dilute (open circles) and dense (filled circles) phases of the specified parent. At intermediate affinity, the parent lies outside the coexistence region and remains homogeneous.
    (B) Fractions of scaffold binding sites participating in scaffold--scaffold ($S$--$S$, blue) and scaffold--client ($S$--$C$, orange) bonds, or remaining unbound (gray), along the affinity scan. The fractions describe the equilibrium dense phase in the phase-separated regimes and the homogeneous parent in the intervening single-phase regime. The homogeneous interval extends approximately from $\beta\epsilon_{SC}=0.5874$ to $3.0729$ and is distinguished by the lighter bond colors. Black triangles mark the five affinities shown in panel A.}
    \label{fig:reentrance}
\end{figure}

We first consider an $A_6+C_3$ mixture at fixed scaffold cohesion, $\beta\epsilon_{SS}=1.6$, and follow the parent composition $(s_0,c_0)=(0.20,0.20)$ as scaffold--client affinity is increased. This parent is chosen to pass through scaffold-rich, homogeneous, and heterotypic regimes as the relative strengths of the molecular interactions are varied. The corresponding coexistence regions and equilibrium tie lines are shown at five representative affinities in Fig.~\ref{fig:reentrance}A.

At weak scaffold--client attraction, the parent lies within a scaffold-rich coexistence region, where demixing is primarily associated with scaffold self-association. Increasing $\epsilon_{SC}$ progressively contracts this region until the parent becomes homogeneous. At stronger heterotypic attraction, a coexistence region supported predominantly by scaffold--client association expands to include the same parent composition, and phase separation reappears (Fig.~\ref{fig:reentrance}A). The resulting affinity response is reentrant, but the two coexistence regimes differ in both their compositions and the molecular associations that favor them. The initial suppression and subsequent recovery of demixing therefore reflect a change in the relative thermodynamic favorability of scaffold-rich and heterotypic association rather than a monotonic strengthening of a single condensed state.

The accompanying redistribution of scaffold binding sites provides a molecular description of this change (Fig.~\ref{fig:reentrance}B). In the scaffold-rich regime, most occupied scaffold sites participate in $S$--$S$ association, and relatively few clients are incorporated into the dense phase. As heterotypic attraction becomes stronger, scaffold sites are increasingly occupied by clients, and $S$--$C$ bonds become the dominant association type in the dense phase. Within the intervening homogeneous regime, the site fractions instead describe the binding state of the homogeneous parent mixture. Thus, increasing client affinity changes not only whether the mixture phase-separates but also how the available scaffold binding sites are allocated. This reentrant behavior motivates a more general question: what determines whether stronger scaffold--client binding increases or decreases the thermodynamic advantage of demixing at a fixed parent composition?

\subsection*{Differential heterotypic bonding sets the local affinity response}

\begin{figure}[!htbp]
    \centering
    \includegraphics[
        width=\linewidth,
        height=.85\textheight,
        keepaspectratio
    ]{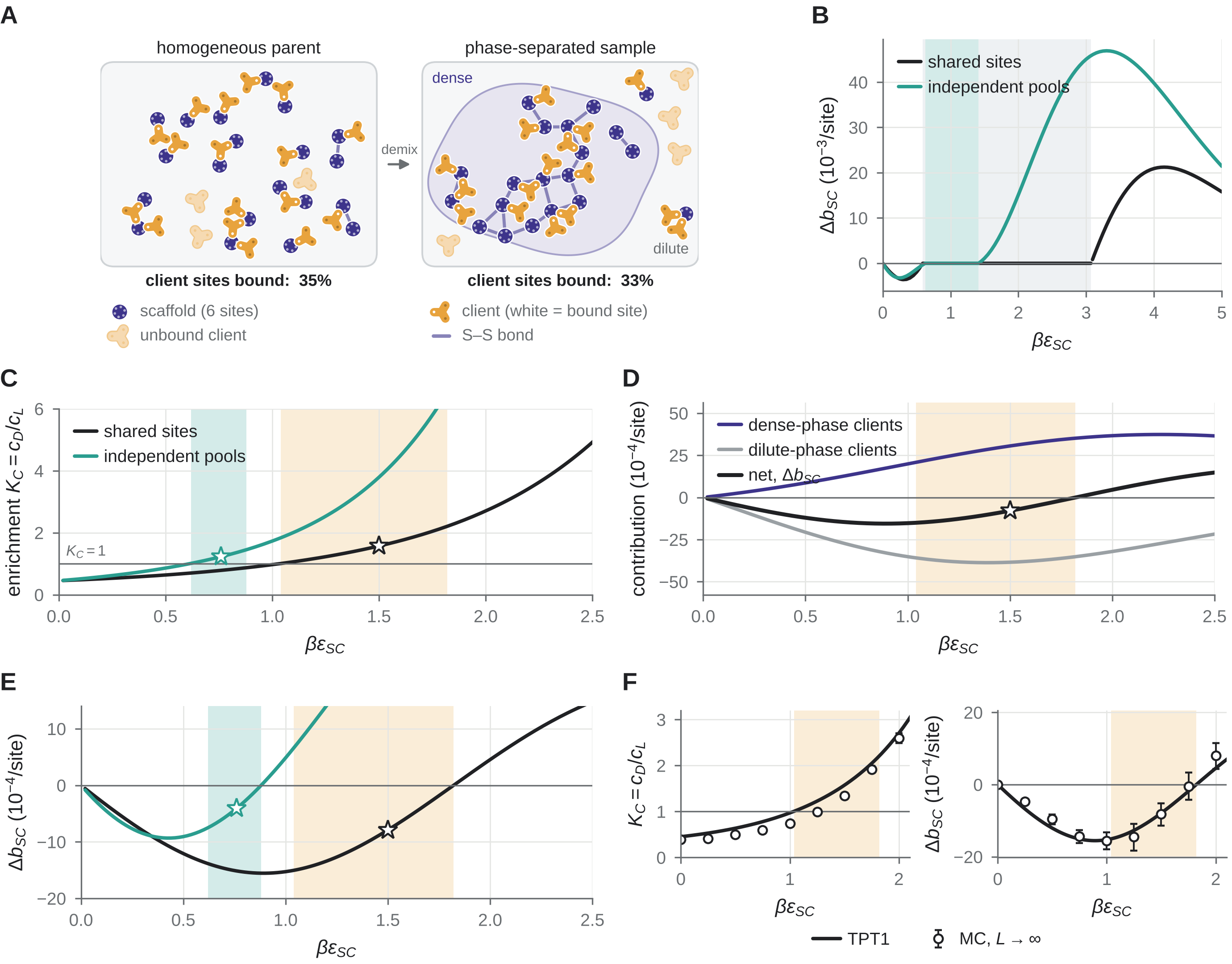}
    \caption{\textbf{Client enrichment does not determine the affinity response of scaffold--client condensation.} (A) Association in homogeneous and phase-separated \(A_6+C_3\) samples, illustrating a decrease in occupied client sites from approximately 35\% to 33\%. (B,E) Heterotypic bond-density contrast, \(\Delta b_{SC}\), for shared-site (black) and independent-pool (green) architectures. (C) Client enrichment factor, \(K_C=c_D/c_L\). (D) Dense- and dilute-phase contributions to the total bond contrast. Shading marks homogeneous intervals in B and co-enriched negative-response intervals in C--E. Stars mark representative states. Panels A--E use \(\beta\epsilon_{SS}=1.6\), with parent composition \((s_0,c_0)=(0.20,0.20)\) in B and \((0.33,0.0195)\) in A,C--E; panel A uses \(\beta\epsilon_{SC}=1.5\). (F) Shared-site TPT1 curves and Monte Carlo estimates at the same nominal parent and scaffold affinity. The left panel compares the bulk partition coefficient (TPT1) with the fitted core enrichment ratio (MC); the right compares heterotypic bond contrasts. MC symbols are weighted \(1/L\) fit intercepts from \(L=8,10,12,16\), with pointwise 95\% bootstrap intervals. The MC bond contrast uses a scaffold-constrained reference. Orange shading denotes the shared-site TPT1 co-enriched negative-response interval.}
    \label{fig:affinity_response}
\end{figure}

A client may accumulate inside a condensate without making stronger scaffold--client binding more favorable to phase separation. The reason is that binding also occurs outside the condensate and in the homogeneous reference. Strengthening the interaction benefits both thermodynamic states. To determine the response, we must therefore compare the same total numbers of scaffolds and clients arranged in two ways: uniformly mixed, or distributed between the equilibrium dilute and dense phases (Fig.~\ref{fig:affinity_response}A). The homogeneous reference is evaluated at the overall, or parent, composition, even when a phase-separated state has lower free energy.

For a fixed parent $(s_0,c_0)$, we quantify the thermodynamic advantage of demixing by $G$, defined in Eq.~\ref{eq:G}. Along the affinity trajectory considered above, $G$ initially decreases, vanishes over the homogeneous interval, and rises again once heterotypic coexistence becomes favorable. Stronger heterotypic binding can therefore either reduce or increase the demixing advantage, depending on the coexistence regime. The origin of this distinction can be expressed directly in terms of molecular association. Differentiating the finite-site free energy at fixed parent composition gives
\begin{equation}
\frac{dG}{d\ln\Delta_{SC}}
=
\langle b_{SC}\rangle_{\mathrm{eq}}
-
b_{SC}^{\mathrm{hom}}
\equiv\Delta b_{SC},
\label{eq:bond_response_results}
\end{equation}
where $\langle b_{SC}\rangle_{\mathrm{eq}}$ is the heterotypic bond density averaged over both equilibrium phases, weighted by their volume fractions. Since $\Delta_{SC}=v[\exp(\beta\epsilon_{SC})-1]$, the interaction-energy derivative is $dG/d(\beta\epsilon_{SC})=\Delta b_{SC}/[1-\exp(-\beta\epsilon_{SC})]$; both derivatives have the same sign for positive binding energies. When $\Delta b_{SC}<0$, the homogeneous parent benefits more from stronger binding, reducing the free-energy advantage of demixing. Conversely, $\Delta b_{SC}>0$ indicates that the phase-separated sample benefits more. The opposing responses are illustrated for shared and independent scaffold binding-site architectures (Fig.~\ref{fig:affinity_response}B).

The distinction between molecular bonding and concentration partitioning becomes apparent by comparison with a bilinear heterotypic interaction. For a reduced free energy $\beta f=\beta f_0-\lambda sc$, with $f_0$ independent of $\lambda$, the corresponding fixed-parent response is
\begin{equation}
\frac{dG}{d\lambda}
=
\langle sc\rangle_{\mathrm{eq}}-s_0c_0
=
\phi(1-\phi)(s_\alpha-s_\gamma)(c_\alpha-c_\gamma)
\equiv\mathcal C_{SC}.
\label{eq:concentration_response}
\end{equation}
For this interaction, the response is determined by the concentration covariance: enrichment of both species in the same phase necessarily increases the demixing advantage as $\lambda$ grows. With finite-site association, the heterotypic bond density depends nonlinearly on composition through site availability and occupancy. Its bond contrast can therefore have a different sign from concentration covariance. Physically, this distinction arises because bringing more molecules together does not necessarily increase their number of bonds proportionally. Once binding sites become occupied, further association is constrained by the finite binding capacity of each molecule.

The trivalent-client system exhibits a finite regime in which these two signs differ: scaffolds and clients both enrich in the dense phase, yet the complete separated sample contains fewer scaffold--client bonds than the homogeneous reference (Fig.~\ref{fig:affinity_response}C,E). Stronger heterotypic binding then reduces the demixing advantage. Thus, the location of a recruited protein and the thermodynamic effect of changing its affinity are distinct properties.

A numerical example makes the molecular origin explicit. At the state illustrated in Fig.~\ref{fig:affinity_response}A, with $(s_0,c_0)\simeq(0.33,0.0195)$ and $\beta\epsilon_{SC}=1.5$, approximately $65.2\%$ of the clients reside in the dense phase, where $42.7\%$ of their binding sites are occupied. The remaining $34.8\%$ reside in the dilute phase, where site occupancy falls to $15.8\%$. Averaged over all clients, the occupied-site fraction is therefore $33.4\%$, below the homogeneous-reference value of $34.7\%$. Clients inside the condensate gain bonds relative to that reference, but this gain is outweighed by the loss among clients outside it.

This accounting can be written for any two-phase state. Let $p_i=1-X_C^i$ denote the fraction of occupied client sites in the homogeneous parent ($i=0$), dilute phase ($i=L$), or dense phase ($i=D$). Since $b_{SC}^i=nc_i p_i$, conservation of clients gives
\begin{equation}
\frac{\Delta b_{SC}}{nc_0}
=
w_D(p_D-p_0)+(1-w_D)(p_L-p_0),
\qquad
w_D=\frac{\phi c_D}{c_0},
\label{eq:occupancy_decomposition}
\end{equation}
where $\phi$ is the dense-phase volume fraction and $w_D$ is the fraction of all clients residing in that phase. The two terms count the occupancy gain or loss among dense- and dilute-phase clients, weighted by their share of the client population. In the co-enriched negative-response regime, $p_D>p_0>p_L$, but the negative dilute-phase contribution exceeds the positive dense-phase contribution (Fig.~\ref{fig:affinity_response}D). The dilute phase therefore helps determine the affinity response even when both components enrich in the condensate. At a representative point, the dense-phase volume fraction and the fraction of scaffold in that phase both increase slightly while $G$ decreases, illustrating that a smaller demixing advantage need not mean less condensed material.

The negative response does not require direct competition for scaffold binding sites. A familiar mechanism of condensate inhibition involves binding partners that interfere with the molecular interactions responsible for scaffold self-association, as illustrated by ubiquitin-dependent regulation of UBQLN2 condensation \cite{dao2018}. However, a regulatory molecule may also interact with sites that are distinct from those responsible for scaffold assembly. To examine this possibility, we consider an architecture in which scaffold self-association and client binding use independent site pools. The negative response persists at weak affinity, while a positive response accompanies heterotypic coexistence (Fig.~\ref{fig:affinity_response}B). At weak affinity, the heterotypic bond density satisfies $b_{SC}\simeq6nsc\Delta_{SC}$, so the bond contrast follows concentration covariance at leading order. Segregative partitioning therefore favors the homogeneous parent by reducing scaffold--client overlap. At finite affinity, nonlinear site occupancy also permits a negative response under co-enrichment, even when scaffold self-association and client binding use independent sites (Fig.~\ref{fig:affinity_response}C,E). These two regimes show that neither direct site displacement nor segregation of the components is necessary for a negative affinity response. Thus, the effect of a binding partner on condensation cannot be understood solely by asking whether it blocks scaffold self-association; it also depends on how its binding is distributed between the competing thermodynamic states.

\subsection*{Alternative molecular associations organize distinct coexistence regimes}

\begin{figure}[!htbp]
    \centering
    \includegraphics[
        width=\linewidth,
        height=.85\textheight,
        keepaspectratio
    ]{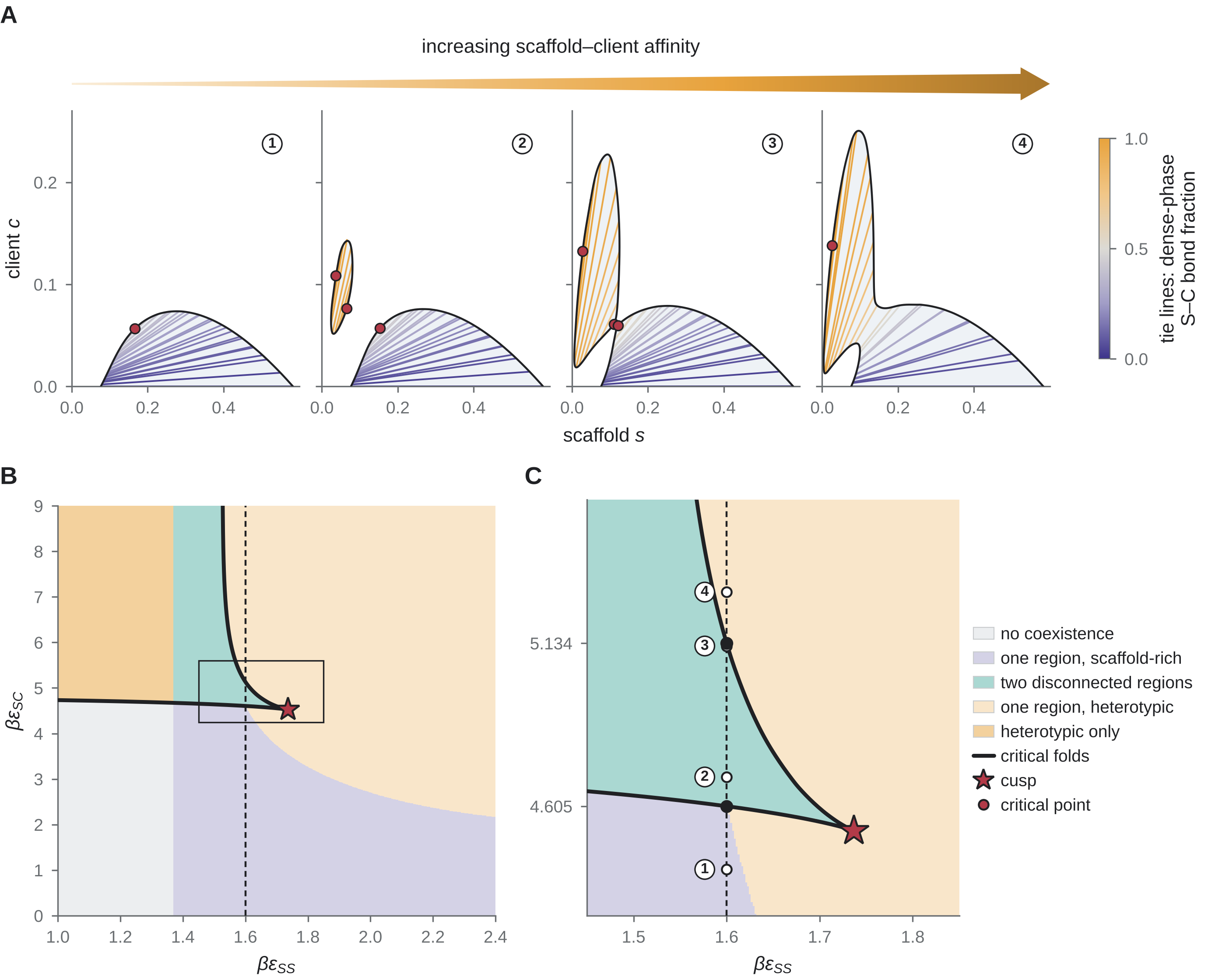}
    \caption{\textbf{Heterotypic coexistence appears as a distinct region and subsequently reconnects with scaffold-rich coexistence as interaction strength increases.} (A) Representative ternary phase diagrams of the $A_6+C_2$ system at $\beta\epsilon_{SS}=1.6$ and increasing scaffold--client affinities $\beta\epsilon_{SC}=4.4$, $4.7$, $5.125$, and $5.3$ (left to right). At weak heterotypic attraction, only scaffold-rich coexistence is present. A second, heterotypic coexistence region subsequently appears at finite client concentration, remains disconnected from the scaffold-rich region, and eventually merges with it. Tie lines connect coexisting dilute and dense compositions and are colored by the heterotypic fraction $q_{SC}$ of the bond population at the dense endpoint. Orange circles denote critical points. (B) Classification of coexistence regimes in the $(\beta\epsilon_{SS},\beta\epsilon_{SC})$ parameter plane. Background colors distinguish the absence of coexistence, a single scaffold-rich or heterotypic coexistence region, two disconnected regions, and heterotypic-only coexistence. Black curves indicate folds of the interior critical-point locus, which meet at a cusp (yellow star). The dashed vertical line marks $\beta\epsilon_{SS}=1.6$, corresponding to panel A. (C) Magnification of the fold and cusp region. Along the reference cut, the heterotypic coexistence region appears at $\beta\epsilon_{SC}=4.605$ and reconnects with the scaffold-rich region at $\beta\epsilon_{SC}=5.134$. Filled black circles mark these thresholds, while numbered open circles correspond to the four interaction strengths illustrated in panel A. The critical-point folds organize changes in the number of interior critical points; the connectivity of the globally stable coexistence regions is determined separately from the equilibrium free energy.}
    \label{fig:topology}
\end{figure}

Biomolecular condensates can be organized through different combinations of homotypic and heterotypic interactions, even when they contain the same molecular components. For example, the balance between NPM1 self-association and NPM1--SURF6 interactions can change with protein composition and molecular crowding \cite{ferrolino2018}. Such observations raise a broader question: can different molecular interaction networks support distinct modes of condensation within the same mixture? To examine this possibility, we turn to the $A_6+C_2$ system, for which scaffold-rich and heterotypic coexistence regimes are more clearly separated. At fixed interaction strengths, two distinct families of coexistence can occupy different regions of the ternary phase diagram (Fig.~\ref{fig:topology}A). One extends from the client-poor side and is dominated by scaffold--scaffold association, while the other occupies a client-richer composition range and contains a substantially larger fraction of scaffold--client bonds.

The two coexistence regimes arise from the same molecular association rules, but differ in how the available scaffold binding sites are used. In the scaffold-rich regime, homotypic interactions dominate the bond population, whereas in the heterotypic regime, a larger fraction of bonds involves clients. The heterotypic bond fraction $q_{SC}$ distinguishes these bonding organizations, as illustrated by the tie-line colors in Fig.~\ref{fig:topology}A. However, molecular partitioning and bonding organization are not equivalent. Tie lines characterize how scaffold and client concentrations change between coexisting phases, whereas bond fractions describe how the available interaction sites are occupied. Consequently, increasingly heterotypic partitioning does not necessarily correspond to a monotonic increase in the occupation of scaffold sites by clients. This distinction is relevant when interpreting biological condensates, where changes in molecular composition may be accompanied by changes in the interactions available for recruiting additional partners or supporting biochemical processes.

The relationship between the two condensation regimes changes as scaffold--client affinity increases. At weak heterotypic attraction, only the scaffold-rich coexistence region is present. Increasing $\epsilon_{SC}$ produces a second coexistence region at finite client concentration, bounded by its own pair of critical points (Fig.~\ref{fig:topology}A). This heterotypic region is initially disconnected from the scaffold-rich region, with stable homogeneous compositions separating the two. Further increases in affinity expand the heterotypic region toward scaffold-richer compositions until the two coexistence regions reconnect. Beyond reconnection, a single connected coexistence region extends across compositions with predominantly homotypic and strongly heterotypic bonding character. Thus, changes in molecular affinity can alter not only the extent of condensation but also the organization and connectivity of the coexistence regions in composition space.

The intervening homogeneous compositions distinguish disconnected coexistence from a continuous crossover in bonding character. For a suitable change in the relative abundance of scaffolds and clients, a mixture may therefore pass from one mode of condensation through a homogeneous regime before entering another. After reconnection, the relative contributions of $S$--$S$ and $S$--$C$ association can instead change smoothly within a connected coexistence region without an additional phase transition. These possibilities provide a physical picture of how condensates composed of the same molecular species could exhibit different dominant interaction networks under different conditions. Bond populations characterize the molecular organization of coexistence, whereas its connectivity is determined by the global free-energy landscape.

Client valence further controls the conditions under which the two coexistence regimes can be separated. For an $A_6+C_n$ mixture, the complementary binding capacities are balanced at $nc=6s$, corresponding to $c/s=6/n$. Increasing client valence shifts this reference composition scale toward scaffold-richer mixtures, but does not by itself determine phase equilibrium or tie-line direction. In the present model, disconnected coexistence occupies a substantial parameter range for divalent clients, becomes more restricted for trivalent clients, and is absent over the investigated ranges for monovalent and tetravalent clients. Monovalent clients cannot bridge scaffolds, whereas changes in multivalent client functionality modify the composition range over which heterotypic association becomes favorable. The existence and separation of the two coexistence regimes therefore depend on the interplay between molecular valence, scaffold cohesion, and heterotypic binding rather than on binding strength alone.

The appearance and reconnection of heterotypic coexistence are organized by the critical-point structure in interaction space. Continuation of the interior critical points in $(\beta\epsilon_{SS},\beta\epsilon_{SC})$ identifies two fold branches that meet at a cusp (Fig.~\ref{fig:topology}B,C). Along the fold locus, the Jacobian of the critical-point conditions becomes singular, marking changes in the number of interior critical points. The folds and cusp organize these changes in critical-point structure; connectivity is assessed independently by branch continuation and global tangent-plane tests. The scans spanning both folds found no competing three-phase state at $\beta\epsilon_{SS}=1.6$. The divalent Monte Carlo sweep resolves separate scaffold-rich and client-rich regions at weaker scaffold cohesion and a connected region at stronger cohesion. This trend supports the predicted dependence on scaffold cohesion, although a narrow connection below the simulation resolution cannot be excluded. Together, these results demonstrate how finite molecular binding constraints can support alternative modes of condensation and reorganize their coexistence as interaction strengths and molecular valence are varied.

\section*{Discussion}

The finite-site framework connects reentrant condensation, fixed-parent affinity response, and coexistence organization through a common free energy and explicit molecular bond populations. Because the binding rules are specified site by site, the same approach can be adapted to other multivalent mixtures by changing site numbers, affinities, and allowed partners. It is most directly applicable when saturable binding captures the main interactions driving condensation. Within this framework, we derive the affinity response of a closed mixture by comparing the complete separated sample with the homogeneous parent at the same overall composition. The derivation uses the general thermodynamic relation between a free-energy derivative and the population coupled to the perturbation \cite{wyman1975,wyman1980}. Polyphasic linkage instead relates phase-specific ligand binding to shifts in scaffold saturation concentration \cite{ruff2021linkage}; the fixed-parent bond contrast includes both coexisting phases, their volume fractions, and the homogeneous reference. The resulting co-enriched negative-response regime illustrates how clients in the dense phase can gain occupied sites while a larger loss among dilute-phase clients reduces the total heterotypic bond population of the separated sample. The dilute phase can therefore determine the sign of the complete-sample affinity response.

Constant-interaction Flory--Huggins-type descriptions provide compact representations of multicomponent coexistence and tie-line organization \cite{deviri2021,farag2023}. For a bilinear heterotypic perturbation, the fixed-parent response is determined by concentration covariance. Co-enrichment therefore necessarily increases the demixing advantage as this attraction strengthens. This description does not track finite binding capacity: additional proteins cannot form new bonds at sites that are already occupied. Finite-site association instead retains bond identity and saturation, so the response is governed by a bond population whose composition dependence follows from mass action. More general nonlinear effective-interaction models can also depart from the bilinear sign rule; the association framework provides a molecular interpretation rather than an exclusive formalism for this behavior. In the Wertheim/SAFT description, bond populations and phase equilibrium follow from the same free energy, allowing distinct coexistence regimes to be examined without assigning separate effective interactions to each. This complements approaches based on coexistence compositions and phase-boundary geometry \cite{qian2024,ausserwoger2025}.

The separate-pool control shows that a negative affinity response need not arise from displacement of scaffold self-association sites. Multivalent regulators can influence condensation through different binding architectures \cite{yoshizawa2018,dao2018,nguemaha2018,ruff2021}. Models of chromatin-binding proteins have used distinct domains for protein self-association and chromatin binding \cite{ancona2022}. Ubiquitin binding to UBQLN2 provides a biological example of interference with the interactions supporting scaffold assembly. In the present model, segregation reduces scaffold--client overlap at weak affinity, whereas nonlinear occupancy permits a negative response under co-enrichment at finite affinity. Neither mechanism requires shared binding sites. The control tests the necessity of direct competition rather than its quantitative contribution, since the independent-pool scaffold has greater total binding capacity than the shared-site architecture.

Disconnected miscibility gaps and changes in sticker allocation have previously been established in associating polymer models \cite{michels2022}. Here, distinct scaffold-rich and heterotypic coexistence regions arise in an equal-size lattice mixture with only saturable $S$--$S$ and $S$--$C$ interactions. Their bond populations distinguish the two organizations, but the heterotypic site-balance scale alone does not determine phase equilibrium \cite{chattaraj2021,deviri2021,farag2023}. Mixing entropy and competing associations allow both regions to be stable at the same interaction parameters while intervening compositions remain homogeneous. At $\beta\epsilon_{SS}=1.5$, a globally stable homogeneous curve separates the two coexistence families. Interconversion between homotypic and heterotypic interactions in NPM1--SURF6 condensates illustrates the biological relevance of alternative bonding organizations \cite{ferrolino2018}, although it does not establish disconnected coexistence in that system. Thus, bond allocation identifies the molecular character of coexistence, while the convexified free energy determines its stability and connectivity.

The separation between scaffold-rich and heterotypic coexistence depends sensitively on molecular valence and scaffold cohesion. Finite valence and binding stoichiometry are already known to generate nonmonotonic phase behavior and suppress macroscopic condensation in associating systems \cite{bianchi2006,delasheras2011,xu2020,zhang2021}. In the present model, disconnected coexistence persists across the tested scaffold valences and cohesion ranges for divalent clients, but occupies a considerably narrower parameter range for trivalent clients and is absent in the investigated monovalent and tetravalent systems. Monovalent clients cannot bridge scaffolds, while increasing multivalent client functionality shifts the heterotypic site-balance scale toward scaffold-rich compositions. This reduces the separation in composition space between the two bonding organizations, although their equilibrium coexistence remains determined by the full free energy. Disconnected coexistence is therefore neither a universal consequence of finite valence nor a monotonic function of client functionality. Scaffold cohesion also affects whether the two regions reconnect as heterotypic affinity increases or remain separated over the investigated range. The relevant molecular parameters influence not only the extent of phase separation but also the arrangement of its distinct coexistence regions.

The folds and cusp of the critical-point locus characterize the parameter regime in which the coexistence topology changes. Near the cusp, the separation between the fold branches follows the generic $3/2$ scaling, with the model-dependent prefactor obtained from free-energy derivatives. These local singularities organize changes in critical-point structure, while global stability determines the connectivity of coexistence regions. The appearance of heterotypic coexistence and its reconnection with scaffold-rich coexistence consequently remain distinct thermodynamic features.

Explicit-particle Monte Carlo simulations provide independent tests of several predictions of the association model. Finite-volume calculations recover the bond-contrast relation between interaction perturbations and free-energy response, including changes in its sign for both shared-site and independent-site architectures. In Fig.~\ref{fig:affinity_response}F, the Monte Carlo bond contrast follows the TPT1 trend and agrees closely at the representative affinity where clients are enriched despite a negative contrast. The comparison uses a scaffold-constrained reference. The scaffold-affinity sweep supports the predicted separation of the coexistence regions at lower scaffold cohesion, with broader simulated coexistence regions than in TPT1 at the same nominal affinities. The comparisons therefore support the principal physical mechanisms while also indicating limitations in the quantitative phase boundaries predicted by the independent-site approximation.

Thermodynamic preference must also be distinguished from condensate abundance. A reduction in $G$ does not determine changes in dense-phase volume, condensed scaffold fraction, or saturation concentration; those observables require phase compositions and phase fractions along a specified preparation path \cite{ruff2021}. Mapping equilibrium coexistence boundaries as scaffold--client affinity varies at fixed overall concentrations provides a direct experimental connection: if the parent leaves the coexistence region, $G$ undergoes a net decrease from a positive value to zero. Contraction of the coexistence region does not fix the local response at compositions that remain inside. Interaction-specific occupancies would test bond allocation, while the complete-sample contrast additionally requires a homogeneous-parent reference. Maintaining a reference that is metastable or unstable may require a constraint, or its free energy may need thermodynamic reconstruction.

To test the predicted affinity response experimentally, one would vary scaffold--client binding strength at fixed overall concentrations while keeping the underlying scaffold--scaffold affinity unchanged. One route is to introduce mutations in the client's scaffold-binding interface while preserving client valence, linker architecture, and solution conditions. Im2--E9 variants provide a precedent for such affinity tuning in synthetic condensate systems \cite{heidenreich2020}. A scaffold that self-associates through one set of sites and binds clients through a separate module would approximate our independent-pool architecture. Alternatively, light-sensitive iLID--SspB binding could control client recruitment \cite{guntas2015}. In OptoMBP, this interaction recruits a solubilizing partner and reversibly dissolves FUS-based condensates \cite{brumbaughreed2024}. In either approach, scaffold--client binding strength should be measured independently and scaffold-only phase behavior checked. For light-controlled binding, the applicability of an equilibrium description would also need to be established. Keeping scaffold--scaffold affinity fixed still allows the number of scaffold--scaffold bonds to change as clients bind. Combined with the phase-boundary and occupancy measurements described above, these perturbations would test whether stronger client binding reduces the demixing advantage despite client enrichment.

Several limitations delimit the molecular interpretation of the present model. First-order Wertheim thermodynamic perturbation theory describes mean site occupancies and bond densities but does not explicitly resolve closed bonding loops, higher-order spatial correlations, percolation, or network connectivity. The lattice reference also neglects molecular-size asymmetry, sequence-dependent site accessibility, conformational coupling, and detailed interaction geometry. These features may matter for biological macromolecules whose interaction sites lie on flexible chains or structured domains. The quantitative differences between TPT1 predictions and explicit-particle Monte Carlo simulations may reflect geometric constraints and bonding correlations absent from the independent-site approximation. In heterotypic condensates, the limited valence of clients restricts the number of connections they can provide between scaffolds, while molecular geometry further constrains simultaneous bond formation. Neglecting these effects may overestimate the tendency toward heterotypic condensation, particularly near its onset. In scaffold-rich condensates, correlations between neighboring bonds and the formation of closed bonding paths may instead provide additional stabilization beyond the independent-site description. These interpretations remain possible explanations rather than separately established mechanisms of the observed deviations. More detailed association theories and simulations that resolve molecular connectivity would be needed to distinguish their contributions.

\section*{Conclusion}

Finite-site association thermodynamics connects molecular bond allocation with affinity response and coexistence organization in scaffold--client mixtures. At fixed parent composition, the response to stronger heterotypic binding is determined by the bond contrast between the homogeneous parent and the complete phase-separated sample. The model identifies a finite region where both species enrich in the dense phase, yet stronger affinity reduces the demixing advantage, including for independent binding-site pools. Across composition space, the same free energy distinguishes scaffold-rich and heterotypic coexistence and describes how their connectivity depends on interaction strengths and molecular valence.

Explicit-particle Monte Carlo simulations support the qualitative bond-response behavior and show separate coexistence regions in the sampled finite-volume landscapes, while exposing quantitative limitations of TPT1. Because interaction-specific bond populations and phase equilibrium follow from the same free energy, molecular binding constraints can be related to both local affinity regulation and global coexistence organization. Bond occupancy and phase behavior therefore provide complementary information beyond molecular recruitment alone.

\FloatBarrier
\setlength{\bibsep}{2pt}
\bibliographystyle{unsrtnat}
\bibliography{references}

\begin{thebibliography}{50}
\providecommand{\natexlab}[1]{#1}
\providecommand{\url}[1]{\texttt{#1}}
\expandafter\ifx\csname urlstyle\endcsname\relax
  \providecommand{\doi}[1]{doi: #1}\else
  \providecommand{\doi}{doi: \begingroup \urlstyle{rm}\Url}\fi

\bibitem[Hyman et~al.(2014)Hyman, Weber, and J\"ulicher]{hyman2014}
Anthony~A. Hyman, Christoph~A. Weber, and Frank J\"ulicher.
\newblock {Liquid-Liquid Phase Separation in Biology}.
\newblock \emph{Annual Review of Cell and Developmental Biology}, 30:\penalty0
  39--58, 2014.
\newblock \doi{10.1146/annurev-cellbio-100913-013325}.

\bibitem[et~al(2017{\natexlab{a}})]{banani2017}
Salman F.~Banani et~al.
\newblock {Biomolecular condensates: organizers of cellular biochemistry}.
\newblock \emph{Nature Reviews Molecular Cell Biology}, 18:\penalty0 285--298,
  2017{\natexlab{a}}.
\newblock \doi{10.1038/nrm.2017.7}.

\bibitem[Choi et~al.(2020)Choi, Holehouse, and Pappu]{choi2020}
Jeong-Mo Choi, Alex~S. Holehouse, and Rohit~V. Pappu.
\newblock {Physical Principles Underlying the Complex Biology of Intracellular
  Phase Transitions}.
\newblock \emph{Annual Review of Biophysics}, 49:\penalty0 107--133, 2020.
\newblock \doi{10.1146/annurev-biophys-121219-081629}.

\bibitem[Lyon et~al.(2021)Lyon, Peeples, and Rosen]{lyon2021}
Andrew~S. Lyon, William~B. Peeples, and Michael~K. Rosen.
\newblock A framework for understanding the functions of biomolecular
  condensates across scales.
\newblock \emph{Nature Reviews Molecular Cell Biology}, 22\penalty0
  (3):\penalty0 215--235, 2021.
\newblock \doi{10.1038/s41580-020-00303-z}.

\bibitem[Li et~al.(2012)Li, Banjade, Cheng, Kim, Chen, Guo, Llaguno,
  Hollingsworth, King, Banani, Russo, Jiang, Nixon, and Rosen]{li2012}
Pilong Li, Sudeep Banjade, Hui-Chun Cheng, Soyeon Kim, Baoyu Chen, Liang Guo,
  Marc Llaguno, Javoris~V. Hollingsworth, David~S. King, Salman~F. Banani,
  Paul~S. Russo, Qiu-Xing Jiang, B.~Tracy Nixon, and Michael~K. Rosen.
\newblock {Phase transitions in the assembly of multivalent signalling
  proteins}.
\newblock \emph{Nature}, 483\penalty0 (7389):\penalty0 336--340, 2012.
\newblock \doi{10.1038/nature10879}.

\bibitem[Banjade and Rosen(2014)]{banjade2014}
Sudeep Banjade and Michael~K. Rosen.
\newblock {Phase transitions of multivalent proteins can promote clustering of
  membrane receptors}.
\newblock \emph{eLife}, 3:\penalty0 e04123, 2014.
\newblock \doi{10.7554/eLife.04123}.

\bibitem[et~al(2017{\natexlab{b}})]{harmon2017}
Tyler S.~Harmon et~al.
\newblock {Intrinsically Disordered Linkers Determine the Interplay between
  Phase Separation and Gelation in Multivalent Proteins}.
\newblock \emph{eLife}, 6:\penalty0 e30294, 2017{\natexlab{b}}.
\newblock \doi{10.7554/eLife.30294}.

\bibitem[et~al(2018{\natexlab{a}})]{wang2018}
Jie~Wang et~al.
\newblock {A Molecular Grammar Governing the Driving Forces for Phase
  Separation of Prion-like RNA Binding Proteins}.
\newblock \emph{Cell}, 174:\penalty0 688--699.e16, 2018{\natexlab{a}}.
\newblock \doi{10.1016/j.cell.2018.06.006}.

\bibitem[et~al(2020{\natexlab{a}})]{martin2020}
Erik W.~Martin et~al.
\newblock {Valence and Patterning of Aromatic Residues Determine the Phase
  Behavior of Prion-Like Domains}.
\newblock \emph{Science}, 367:\penalty0 694--699, 2020{\natexlab{a}}.
\newblock \doi{10.1126/science.aaw8653}.

\bibitem[et~al(2020{\natexlab{b}})]{riback2020}
Joshua A.~Riback et~al.
\newblock {Composition-Dependent Thermodynamics of Intracellular Phase
  Separation}.
\newblock \emph{Nature}, 581:\penalty0 209--214, 2020{\natexlab{b}}.
\newblock \doi{10.1038/s41586-020-2256-2}.

\bibitem[et~al(2023)]{farag2023}
Mina~Farag et~al.
\newblock {Phase Separation of Protein Mixtures Is Driven by the Interplay of
  Homotypic and Heterotypic Interactions}.
\newblock \emph{Nature Communications}, 14:\penalty0 5527, 2023.
\newblock \doi{10.1038/s41467-023-41274-x}.

\bibitem[et~al(2020{\natexlab{c}})]{sanders2020}
David W.~Sanders et~al.
\newblock {Competing Protein--RNA Interaction Networks Control Multiphase
  Intracellular Organization}.
\newblock \emph{Cell}, 181:\penalty0 306--324.e28, 2020{\natexlab{c}}.
\newblock \doi{10.1016/j.cell.2020.03.050}.

\bibitem[et~al(2018{\natexlab{b}})]{yoshizawa2018}
Takuya~Yoshizawa et~al.
\newblock {Nuclear Import Receptor Inhibits Phase Separation of FUS through
  Binding to Multiple Sites}.
\newblock \emph{Cell}, 173:\penalty0 693--705.e22, 2018{\natexlab{b}}.
\newblock \doi{10.1016/j.cell.2018.03.003}.

\bibitem[et~al(2018{\natexlab{c}})]{dao2018}
Thuy P.~Dao et~al.
\newblock {Ubiquitin Modulates Liquid-Liquid Phase Separation of UBQLN2 via
  Disruption of Multivalent Interactions}.
\newblock \emph{Molecular Cell}, 69:\penalty0 965--978.e6, 2018{\natexlab{c}}.
\newblock \doi{10.1016/j.molcel.2018.02.004}.

\bibitem[et~al(2017{\natexlab{c}})]{banerjee2017}
Priya R.~Banerjee et~al.
\newblock {Reentrant Phase Transition Drives Dynamic Substructure Formation in
  Ribonucleoprotein Droplets}.
\newblock \emph{Angewandte Chemie International Edition}, 56:\penalty0
  11354--11359, 2017{\natexlab{c}}.
\newblock \doi{10.1002/anie.201703191}.

\bibitem[et~al(2018{\natexlab{d}})]{maharana2018}
Shovamayee~Maharana et~al.
\newblock {RNA Buffers the Phase Separation Behavior of Prion-Like RNA Binding
  Proteins}.
\newblock \emph{Science}, 360:\penalty0 918--921, 2018{\natexlab{d}}.
\newblock \doi{10.1126/science.aar7366}.

\bibitem[et~al(2019)]{alshareedah2019}
Ibraheem~Alshareedah et~al.
\newblock {Interplay between Short-Range Attraction and Long-Range Repulsion
  Controls Reentrant Liquid Condensation of Ribonucleoprotein--RNA Complexes}.
\newblock \emph{Journal of the American Chemical Society}, 141:\penalty0
  14593--14602, 2019.
\newblock \doi{10.1021/jacs.9b03689}.

\bibitem[Sanchez-Burgos et~al.(2023)Sanchez-Burgos, Herriott,
  Collepardo-Guevara, and Espinosa]{sanchezburgos2023}
Ignacio Sanchez-Burgos, Lara Herriott, Rosana Collepardo-Guevara, and Jorge~R.
  Espinosa.
\newblock Surfactants or scaffolds? {RNAs} of varying lengths control the
  thermodynamic stability of condensates differently.
\newblock \emph{Biophysical Journal}, 122\penalty0 (14):\penalty0 2973--2987,
  2023.
\newblock \doi{10.1016/j.bpj.2023.03.006}.

\bibitem[Nguemaha and Zhou(2018)]{nguemaha2018}
Vincent Nguemaha and Huan-Xiang Zhou.
\newblock {Liquid--liquid phase separation of patchy particles illuminates
  diverse effects of regulatory components on protein droplet formation}.
\newblock \emph{Scientific Reports}, 8:\penalty0 6728, 2018.
\newblock \doi{10.1038/s41598-018-25132-1}.

\bibitem[Ghosh et~al.(2019)Ghosh, Mazarakos, and Zhou]{ghosh2019}
Amarnath Ghosh, Katerina Mazarakos, and Huan-Xiang Zhou.
\newblock {Three Archetypical Classes of Macromolecular Regulators of Protein
  Liquid--Liquid Phase Separation}.
\newblock \emph{Proceedings of the National Academy of Sciences of the United
  States of America}, 116:\penalty0 19474--19483, 2019.
\newblock \doi{10.1073/pnas.1907849116}.

\bibitem[et~al(2020{\natexlab{d}})]{espinosa2020}
Jorge R.~Espinosa et~al.
\newblock {Liquid Network Connectivity Regulates the Stability and Composition
  of Biomolecular Condensates with Many Components}.
\newblock \emph{Proceedings of the National Academy of Sciences of the United
  States of America}, 117:\penalty0 13238--13247, 2020{\natexlab{d}}.
\newblock \doi{10.1073/pnas.1917569117}.

\bibitem[Joseph et~al.(2021)Joseph, Espinosa, Sanchez-Burgos, Garaizar,
  Frenkel, and Collepardo-Guevara]{joseph2021}
Jerelle~A. Joseph, Jorge~R. Espinosa, Ignacio Sanchez-Burgos, Adiran Garaizar,
  Daan Frenkel, and Rosana Collepardo-Guevara.
\newblock Thermodynamics and kinetics of phase separation of protein-{RNA}
  mixtures by a minimal model.
\newblock \emph{Biophysical Journal}, 120\penalty0 (7):\penalty0 1219--1230,
  2021.
\newblock \doi{10.1016/j.bpj.2021.01.031}.

\bibitem[Ruff et~al.(2021{\natexlab{a}})Ruff, Dar, and Pappu]{ruff2021}
Kiersten~M. Ruff, Furqan Dar, and Rohit~V. Pappu.
\newblock {Ligand Effects on Phase Separation of Multivalent Macromolecules}.
\newblock \emph{Proceedings of the National Academy of Sciences of the United
  States of America}, 118:\penalty0 e2017184118, 2021{\natexlab{a}}.
\newblock \doi{10.1073/pnas.2017184118}.

\bibitem[Ferrolino et~al.(2018)Ferrolino, Mitrea, Michael, and
  Kriwacki]{ferrolino2018}
Mylene~C. Ferrolino, Diana~M. Mitrea, J.~Robert Michael, and Richard~W.
  Kriwacki.
\newblock Compositional adaptability in {NPM1-SURF6} scaffolding networks
  enabled by dynamic switching of phase separation mechanisms.
\newblock \emph{Nature Communications}, 9\penalty0 (1):\penalty0 5064, 2018.
\newblock \doi{10.1038/s41467-018-07530-1}.

\bibitem[Wyman(1975)]{wyman1975}
Jeffries Wyman.
\newblock {A group of thermodynamic potentials applicable to ligand binding by
  a polyfunctional macromolecule}.
\newblock \emph{Proceedings of the National Academy of Sciences of the United
  States of America}, 72:\penalty0 1464--1468, 1975.
\newblock \doi{10.1073/pnas.72.4.1464}.

\bibitem[Wyman and Gill(1980)]{wyman1980}
Jeffries Wyman and Stanley~J. Gill.
\newblock {Ligand-linked phase changes in a biological system: applications to
  sickle cell hemoglobin}.
\newblock \emph{Proceedings of the National Academy of Sciences USA},
  77:\penalty0 5239--5242, 1980.
\newblock \doi{10.1073/pnas.77.9.5239}.

\bibitem[Ruff et~al.(2021{\natexlab{b}})Ruff, Dar, and Pappu]{ruff2021linkage}
Kiersten~M. Ruff, Furqan Dar, and Rohit~V. Pappu.
\newblock {Polyphasic linkage and the impact of ligand binding on the
  regulation of biomolecular condensates}.
\newblock \emph{Biophysics Reviews}, 2:\penalty0 021302, 2021{\natexlab{b}}.
\newblock \doi{10.1063/5.0050059}.

\bibitem[Banani et~al.(2016)Banani, Rice, Peeples, Lin, Jain, Parker, and
  Rosen]{banani2016}
Salman~F. Banani, Allyson~M. Rice, William~B. Peeples, Yuan Lin, Saumya Jain,
  Roy Parker, and Michael~K. Rosen.
\newblock Compositional control of phase-separated cellular bodies.
\newblock \emph{Cell}, 166\penalty0 (3):\penalty0 651--663, 2016.
\newblock \doi{10.1016/j.cell.2016.06.010}.

\bibitem[Deviri and Safran(2021)]{deviri2021}
Dan Deviri and Samuel~A. Safran.
\newblock {Physical Theory of Biological Noise Buffering by Multicomponent
  Phase Separation}.
\newblock \emph{Proceedings of the National Academy of Sciences of the United
  States of America}, 118:\penalty0 e2100099118, 2021.
\newblock \doi{10.1073/pnas.2100099118}.

\bibitem[Chattaraj et~al.(2021)Chattaraj, Blinov, and Loew]{chattaraj2021}
Aniruddha Chattaraj, Michael~L. Blinov, and Leslie~M. Loew.
\newblock {The solubility product extends the buffering concept to heterotypic
  biomolecular condensates}.
\newblock \emph{eLife}, 10:\penalty0 e67176, 2021.
\newblock \doi{10.7554/eLife.67176}.

\bibitem[Chattaraj and Loew(2023)]{chattaraj2023}
Aniruddha Chattaraj and Leslie~M. Loew.
\newblock The maximum solubility product marks the threshold for condensation
  of multivalent biomolecules.
\newblock \emph{Biophysical Journal}, 122\penalty0 (9):\penalty0 1678--1690,
  2023.
\newblock \doi{10.1016/j.bpj.2023.03.036}.

\bibitem[et~al(2024)]{qian2024}
Daoyuan~Qian et~al.
\newblock {Dominance Analysis to Assess Solute Contributions to Multicomponent
  Phase Equilibria}.
\newblock \emph{Proceedings of the National Academy of Sciences of the United
  States of America}, 121:\penalty0 e2407453121, 2024.
\newblock \doi{10.1073/pnas.2407453121}.

\bibitem[Ausserw\"oger et~al.(2025)]{ausserwoger2025}
Hannes Ausserw\"oger et~al.
\newblock {Quantifying collective interactions in biomolecular phase
  separation}.
\newblock \emph{Nature Communications}, 16:\penalty0 7724, 2025.
\newblock \doi{10.1038/s41467-025-62437-y}.

\bibitem[Wertheim(1984{\natexlab{a}})]{wertheim1}
Michael~S. Wertheim.
\newblock {Fluids with Highly Directional Attractive Forces. I. Statistical
  Thermodynamics}.
\newblock \emph{Journal of Statistical Physics}, 35:\penalty0 19--34,
  1984{\natexlab{a}}.
\newblock \doi{10.1007/BF01017362}.

\bibitem[Wertheim(1984{\natexlab{b}})]{wertheim2}
Michael~S. Wertheim.
\newblock {Fluids with Highly Directional Attractive Forces. II. Thermodynamic
  Perturbation Theory and Integral Equations}.
\newblock \emph{Journal of Statistical Physics}, 35:\penalty0 35--47,
  1984{\natexlab{b}}.
\newblock \doi{10.1007/BF01017363}.

\bibitem[Wertheim(1986)]{wertheim3}
Michael~S. Wertheim.
\newblock {Fluids with highly directional attractive forces. III. Multiple
  attraction sites}.
\newblock \emph{Journal of Statistical Physics}, 42:\penalty0 459--476, 1986.
\newblock \doi{10.1007/BF01127721}.

\bibitem[et~al(1989)]{chapman1989}
Walter G.~Chapman et~al.
\newblock {SAFT: Equation-of-State Solution Model for Associating Fluids}.
\newblock \emph{Fluid Phase Equilibria}, 52:\penalty0 31--38, 1989.
\newblock \doi{10.1016/0378-3812(89)80308-5}.

\bibitem[Bianchi et~al.(2006)Bianchi, Largo, Tartaglia, Zaccarelli, and
  Sciortino]{bianchi2006}
Emanuela Bianchi, Julio Largo, Piero Tartaglia, Emanuela Zaccarelli, and
  Francesco Sciortino.
\newblock {Phase diagram of patchy colloids: towards empty liquids}.
\newblock \emph{Physical Review Letters}, 97:\penalty0 168301, 2006.
\newblock \doi{10.1103/PhysRevLett.97.168301}.

\bibitem[de~las Heras et~al.(2011)de~las Heras, Tavares, and
  da~Gama]{delasheras2011}
Daniel de~las Heras, Jos\'e~M. Tavares, and Margarida M.~Telo da~Gama.
\newblock {Phase diagrams of binary mixtures of patchy colloids with distinct
  numbers and types of patches: the empty fluid regime}.
\newblock \emph{The Journal of Chemical Physics}, 134:\penalty0 104904, 2011.
\newblock \doi{10.1063/1.3561396}.

\bibitem[Michels et~al.(2022)]{michels2022}
Jasper~J. Michels et~al.
\newblock {Role of solvent compatibility in the phase behavior of binary
  solutions of weakly associating multivalent polymers}.
\newblock \emph{Biomacromolecules}, 23:\penalty0 349--364, 2022.
\newblock \doi{10.1021/acs.biomac.1c01301}.

\bibitem[Xu et~al.(2020)Xu, He, Weiner, Ronceray, Meir, Jonikas, and
  Wingreen]{xu2020}
Bin Xu, Guanhua He, Benjamin~G. Weiner, Pierre Ronceray, Yigal Meir, Martin~C.
  Jonikas, and Ned~S. Wingreen.
\newblock {Rigidity enhances a magic-number effect in polymer phase
  separation}.
\newblock \emph{Nature Communications}, 11\penalty0 (1):\penalty0 1561, 2020.
\newblock \doi{10.1038/s41467-020-15395-6}.

\bibitem[Zhang et~al.(2021)Zhang, Xu, Weiner, Meir, and Wingreen]{zhang2021}
Yaojun Zhang, Bin Xu, Benjamin~G. Weiner, Yigal Meir, and Ned~S. Wingreen.
\newblock {Decoding the physical principles of two-component biomolecular phase
  separation}.
\newblock \emph{eLife}, 10:\penalty0 e62403, 2021.
\newblock \doi{10.7554/eLife.62403}.

\bibitem[Ancona and Brackley(2022)]{ancona2022}
Marco Ancona and Chris~A. Brackley.
\newblock Simulating the chromatin-mediated phase separation of model proteins
  with multiple domains.
\newblock \emph{Biophysical Journal}, 121\penalty0 (13):\penalty0 2600--2612,
  2022.
\newblock \doi{10.1016/j.bpj.2022.05.039}.

\bibitem[Heidenreich et~al.(2020)Heidenreich, Georgeson, Locatelli, Rovigatti,
  Nandi, Steinberg, Nadav, Shimoni, Safran, Doye, and Levy]{heidenreich2020}
Meta Heidenreich, Joseph~M. Georgeson, Emanuele Locatelli, Lorenzo Rovigatti,
  Saroj~Kumar Nandi, Avital Steinberg, Yotam Nadav, Eyal Shimoni, Samuel~A.
  Safran, Jonathan P.~K. Doye, and Emmanuel~D. Levy.
\newblock Designer protein assemblies with tunable phase diagrams in living
  cells.
\newblock \emph{Nature Chemical Biology}, 16:\penalty0 939--945, 2020.
\newblock \doi{10.1038/s41589-020-0576-z}.

\bibitem[Guntas et~al.(2015)Guntas, Hallett, Zimmerman, Williams, Yumerefendi,
  Bear, and Kuhlman]{guntas2015}
Gurkan Guntas, Ryan~A. Hallett, Seth~P. Zimmerman, Tishan Williams, Hayretin
  Yumerefendi, James~E. Bear, and Brian Kuhlman.
\newblock Engineering an improved light-induced dimer ({iLID}) for controlling
  the localization and activity of signaling proteins.
\newblock \emph{Proceedings of the National Academy of Sciences of the United
  States of America}, 112\penalty0 (1):\penalty0 112--117, 2015.
\newblock \doi{10.1073/pnas.1417910112}.

\bibitem[Brumbaugh-Reed et~al.(2024)Brumbaugh-Reed, Gao, Aoki, and
  Toettcher]{brumbaughreed2024}
Ellen~H. Brumbaugh-Reed, Yang Gao, Kazuhiro Aoki, and Jared~E. Toettcher.
\newblock Rapid and reversible dissolution of biomolecular condensates using
  light-controlled recruitment of a solubility tag.
\newblock \emph{Nature Communications}, 15:\penalty0 6717, 2024.
\newblock \doi{10.1038/s41467-024-50858-0}.

\bibitem[Torrie and Valleau(1977)]{torrie1977}
G.~M. Torrie and J.~P. Valleau.
\newblock Nonphysical sampling distributions in {Monte Carlo} free-energy
  estimation: umbrella sampling.
\newblock \emph{Journal of Computational Physics}, 23:\penalty0 187--199, 1977.

\bibitem[Ferrenberg and Swendsen(1989)]{ferrenberg1989}
A.~M. Ferrenberg and R.~H. Swendsen.
\newblock Optimized {Monte Carlo} data analysis.
\newblock \emph{Physical Review Letters}, 63:\penalty0 1195--1198, 1989.

\bibitem[Kumar et~al.(1992)Kumar, Rosenberg, Bouzida, Swendsen, and
  Kollman]{kumar1992}
S.~Kumar, J.~M. Rosenberg, D.~Bouzida, R.~H. Swendsen, and P.~A. Kollman.
\newblock The weighted histogram analysis method for free-energy calculations
  on biomolecules. {I}. the method.
\newblock \emph{Journal of Computational Chemistry}, 13:\penalty0 1011--1021,
  1992.

\bibitem[Borgs and Koteck{\'y}(1990)]{borgs1990}
C.~Borgs and R.~Koteck{\'y}.
\newblock A rigorous theory of finite-size scaling at first-order phase
  transitions.
\newblock \emph{Journal of Statistical Physics}, 61:\penalty0 79--119, 1990.

\end{thebibliography}

\clearpage
\appendix
\setcounter{equation}{0}
\setcounter{figure}{0}
\setcounter{table}{0}
\renewcommand{\thesection}{S\arabic{section}}
\renewcommand{\theequation}{S\arabic{equation}}
\renewcommand{\thefigure}{S\arabic{figure}}
\renewcommand{\thetable}{S\arabic{table}}
\renewcommand{\theHequation}{supplement.\arabic{equation}}
\renewcommand{\theHfigure}{supplement.\arabic{figure}}
\renewcommand{\theHtable}{supplement.\arabic{table}}

\section*{Supporting Material}
\renewcommand{\contentsname}{Contents}
\setlength{\cftsecnumwidth}{2.8em}
\setlength{\cftsubsecnumwidth}{3.6em}
\setlength{\cftbeforesecskip}{4pt}
\tableofcontents
\clearpage
\setlength{\parskip}{3pt}
\section{Model conventions and numerical methods}
\label{si:model}

\subsection{Conventions}
We use the lattice mixture from the main-text Methods. The fractions of sites occupied by scaffolds, clients, and solvent are $s=\rho_S$, $c=\rho_C$, and $z=1-s-c$. In the reference model, a scaffold has six binding sites and a client has $n$. Each site can form one bond, and clients bind scaffolds. We write $X_S$ and $X_C$ for the unbound fractions.

We use first-order Wertheim association theory (TPT1) \cite{wertheim1,wertheim2,chapman1989}. In this theory the energy and bonding volume enter together as $\Delta_{ij}=v[\exp(\beta\epsilon_{ij})-1]$. Holding $\Delta_{ij}$ fixed while changing $v$ lets us compare the same free energy on different energy axes. For a new bonding volume $v'$, the corresponding energy is
\begin{equation}
\beta\epsilon'_{ij}=\ln\!\left[1+\frac{v}{v'}\left(e^{\beta\epsilon_{ij}}-1\right)\right],
\label{eq:S-remap}
\end{equation}
We use $v=1/8$ unless specified otherwise. It gives the same dilute two-particle association weight as the body-centred cubic lattice simulations (Sec.~\ref{si:mc-model}). The reference scaffold cohesion is $\beta\epsilon_{SS}=1.6$.

\subsection{Mass action, thermodynamic functions and derivatives}
At each composition, we first need to know how many sites are available for bonding. Eliminating $X_C$ reduces the two mass-action equations to one. We solve for the physical root $X_S$ by bisection on the interval of possible unbound fractions, then obtain $X_C$ from the client equation.
\begin{equation}
ad\,X_S^3+(a+d)X_S^2+(1+b-d)X_S-1=0,\qquad a=6s\Delta_{SS},\quad b=nc\Delta_{SC},\quad d=6s\Delta_{SC}.
\end{equation}
To locate critical points we need the curvature of the free energy and its higher derivatives. We differentiate the root condition implicitly so these derivatives include the change in site occupancy with composition (Sec.~\ref{si:critical-points}).

\FloatBarrier
\section{Affinity response, client occupancy, and co-enrichment}
\label{si:affinity-response}
\subsection{Analytical basis of the bond-response identity}
\label{si:bond-response-derivation}
To find the affinity response, first ask how changing an interaction affects a uniform mixture at fixed composition. The number of sites stays fixed, but their allocation among free and bound states changes. Write $a=6s$ and $d=nc$ for the site densities. An $S$--$S$ bond consumes two scaffold sites, whereas an $S$--$C$ bond consumes one site of each kind. This gives the site balances and bond densities
\begin{align}
a(1-X_S)&=2b_{SS}+b_{SC},& d(1-X_C)&=b_{SC},\\
b_{SS}&=\tfrac12a^2X_S^2\Delta_{SS},& b_{SC}&=adX_SX_C\Delta_{SC}.
\end{align}
Using the site balances, we can express the association free energy in terms of the unbound fractions and the bond counts.
\begin{equation}
\beta f_{\mathrm{assoc}}=a\ln X_S+d\ln X_C+b_{SS}+b_{SC}.
\end{equation}
At fixed $a$ and $d$, each additional bond must be matched by a loss of unbound sites. The differential form of that balance is
\begin{equation}
aX_S\,d\ln X_S=-2\,db_{SS}-db_{SC},\qquad
dX_C\,d\ln X_C=-db_{SC}.
\end{equation}
Combining this balance with the logarithmic differentials of the bond densities gives
\begin{align}
a\,d\ln X_S+d\,d\ln X_C
={}&-d(b_{SS}+b_{SC})\notag\\
&-b_{SS}\,d\ln\Delta_{SS}-b_{SC}\,d\ln\Delta_{SC}.
\end{align}
The terms from changing the total bond count cancel when we differentiate the free energy. At fixed composition the mixing term is constant, leaving
\begin{equation}
d(\beta f)=-b_{SS}\,d\ln\Delta_{SS}-b_{SC}\,d\ln\Delta_{SC}
\quad\text{at fixed }s,c.
\end{equation}
The response of a uniform phase to $\ln\Delta_{SC}$ is therefore set by the $S$--$C$ bond density it already has. For a sample that separates, we must account for every phase. The equilibrium free-energy density is
\begin{equation}
\beta f_{\mathrm{eq}}=\min_{\{\phi_j,s_j,c_j\}}
\sum_j\phi_j\beta f(s_j,c_j),
\end{equation}
Here $\phi_j$ is the volume fraction of phase $j$. The constraints $\sum_j\phi_j=1$, $\sum_j\phi_js_j=s_0$, and $\sum_j\phi_jc_j=c_0$ keep the total composition fixed. The phase compositions and volumes can change with affinity. Along a smooth equilibrium branch, their first-order changes cancel at the constrained minimum, so only the explicit affinity derivative contributes.
\begin{equation}
\frac{d(\beta f_{\mathrm{eq}})}{d\ln\Delta_{SC}}
=-\sum_j\phi_j b_{SC}^{j}.
\end{equation}
Subtracting the equilibrium derivative from the uniform-parent derivative gives the bond contrast in the main text. This argument also applies when more than two phases minimize the free energy. If the minimizing branch changes, take the derivatives from each side separately.

\subsection{Client-resolved occupancy balance and the sign of the response}
\label{si:occupancy-balance}
To see how the dilute phase can influence the sign, count occupied client sites in each phase. Let $D$ and $L$ denote the dense and dilute phases, $0$ the homogeneous parent, and $\phi$ the dense-phase volume fraction. Each occupied client site corresponds to one heterotypic bond, so $b_{SC}^i=nc_i p_i$ with $p_i=1-X_C^i$. The lever rule, $c_0=\phi c_D+(1-\phi)c_L$, then gives
\begin{align}
\Delta b_{SC}
&=n\left[\phi c_D(p_D-p_0)+(1-\phi)c_L(p_L-p_0)\right],\notag\\
\frac{\Delta b_{SC}}{nc_0}
&=w_D(p_D-p_0)+(1-w_D)(p_L-p_0),
\qquad w_D\equiv\frac{\phi c_D}{c_0}.
\label{eq:S-occupancy-balance}
\end{align}
The weight $w_D$ is the share of all clients in the dense phase. When $p_D>p_0>p_L$, a dense-phase client gains occupied sites and a dilute-phase client loses them, relative to the uniform parent. The losses dominate the sample-wide response when
\begin{equation}
(1-w_D)(p_0-p_L)>w_D(p_D-p_0).
\label{eq:S-occupancy-sign}
\end{equation}
We use a bilinear attraction to see what changes if the interaction depends directly on the two concentrations. For $\beta f=\beta f_0(s,c)-\lambda sc$, with $\beta f_0$ independent of $\lambda$, the same fixed-composition derivative gives
\begin{align}
\frac{dG}{d\lambda}
&=\phi s_Dc_D+(1-\phi)s_Lc_L-s_0c_0\notag\\
&=\phi(1-\phi)(s_D-s_L)(c_D-c_L).
\label{eq:S-bilinear-control}
\end{align}
If both components enrich in the dense phase, both concentration differences are positive. This bilinear perturbation therefore increases the demixing advantage. With finite binding sites, the response depends on occupied sites and their distribution across the whole sample, as in Eq.~\eqref{eq:S-occupancy-sign}.

\subsection{Trivalent-client co-enrichment and the sampled negative-response region}
\label{si:coenrichment}
We first vary affinity at the parent used for the trivalent-client example in main-text Fig.~3, $(s_0,c_0)\simeq(0.33,0.0195)$ with $\beta\epsilon_{SS}=1.6$. This keeps the total amounts fixed, so the scan isolates how affinity changes the balance of occupied sites. At the sampled points, co-enrichment and a negative bond contrast occur over $\beta\epsilon_{SC}=1.04$--$1.82$ for shared sites and $0.62$--$0.88$ for independent pools.

\begin{figure}[htbp]
\centering
\includegraphics[width=\linewidth]{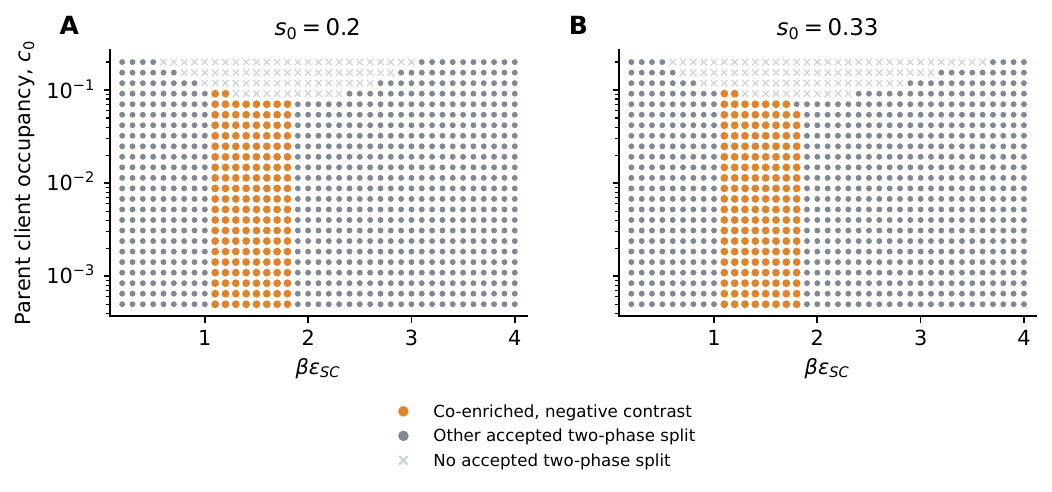}
\caption{Where co-enrichment accompanies a negative affinity response for trivalent clients. In the shared-site model at $\beta\epsilon_{SS}=1.6$, orange points mark two-phase splits with $\mathrm{Cov}(s,c)>0$ and $\Delta b_{SC}<0$. Gray points show the other accepted splits. Pale crosses mark compositions without an accepted split. Panels A and B use $s_0=0.20$ and $0.33$, respectively.}
\label{fig:S-c3-region}
\end{figure}

We then vary both client concentration and affinity for the shared-site model to see whether the effect extends beyond the example. At each grid point we estimate the equilibrium split from the lower convex envelope and refine the endpoints by equating chemical potentials and osmotic pressure. Figure~\ref{fig:S-c3-region} marks the splits that meet both criteria. The effect spans a range of client concentrations at both scaffold concentrations shown, with the strongest sampled contrasts near $c_0=0.05$.

\FloatBarrier
\section{Tie lines and bond allocation across client valence}
\label{si:valence-ties}

\begin{figure}[htbp]
\centering
\includegraphics[width=\linewidth]{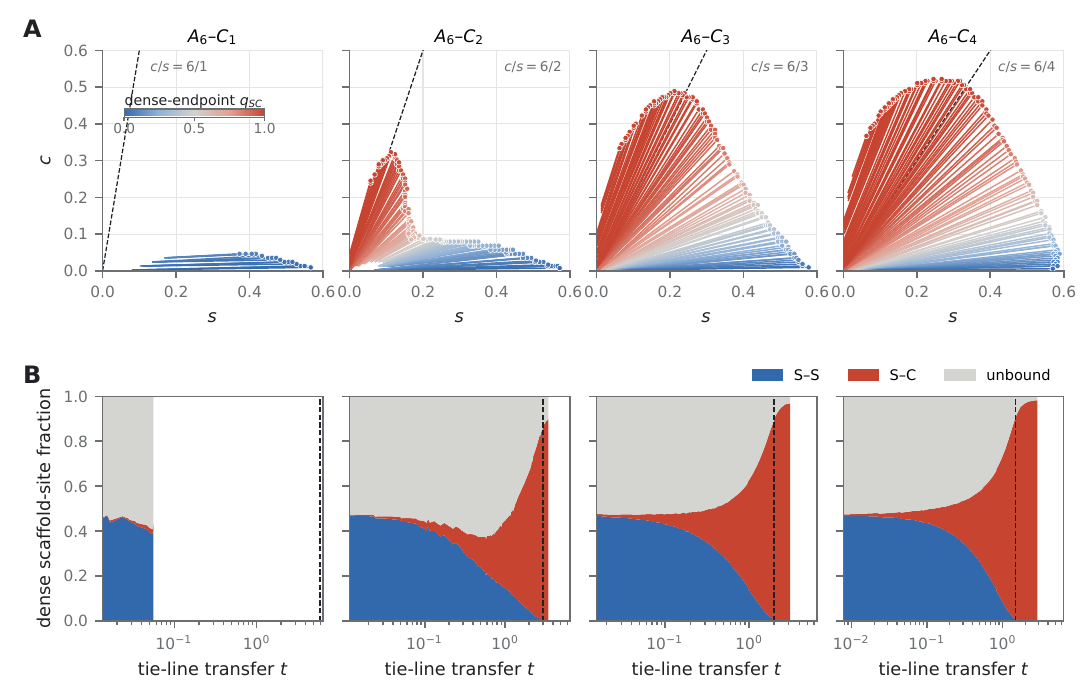}
\caption{Tie lines and dense-phase scaffold-site allocation for $C_1$--$C_4$. $A_6$--$C_n$, $\beta\epsilon_{SS}=1.6$,
$\beta\epsilon_{SC}=6$. (A) Coexisting phases; each tie line is colored by the heterotypic bond fraction $q_{SC}$ of its dense endpoint (dots).
Dashed lines mark the site-balance ratio $c/s=6/n$. (B) Fraction of dense-phase scaffold sites in $S$--$S$ bonds, $S$--$C$ bonds, or
unbound, against tie-line transfer $t$ (logarithmic axis). Dashed lines mark $t=6/n$.}
\label{fig:S-valence-ties}
\end{figure}

A tie line tells us how scaffold and client concentrations change between the two phases. The site occupancies tell us which bonds those molecules form. We pair the two measurements in Fig.~\ref{fig:S-valence-ties} to see how adding client sites changes the dense phase. We take the coexistence endpoints from the lower convex envelope and compare the bond fractions with the tie-line slope $t=\Delta c/\Delta s$.

A monovalent client can bind a scaffold but cannot connect two of them. Its tie lines are almost parallel to the scaffold axis, and only a small fraction of dense-phase scaffold sites bind clients. For multivalent clients, the fraction of $S$--$C$ bonds increases along the steeper tie lines. As valence rises, $S$--$C$ bonds can dominate even at larger scaffold concentrations. This brings the heterotypic states closer in composition to the scaffold-rich states, consistent with the narrower disconnection window at larger valence (Sec.~\ref{si:robustness}). At the affinity shown, the divalent system is above the reconnection threshold, so both kinds of tie line belong to one coexistence region.

\FloatBarrier
\section{Global stability and disconnected coexistence}
\label{si:three-phase}

Could a scaffold-rich pair and a heterotypic pair be part of the same three-phase equilibrium? All three phases would have to share a client chemical potential. We hold $\mu_C$ fixed and minimize over $c$ to test this directly. With $\Delta_{CC}=0$, the free energy is strictly convex in $c$ at fixed $s$, so this step gives one client concentration for each scaffold concentration. The resulting potential is
\begin{equation}
g(s;\mu_C)=\min_c\left[\beta f(s,c)-\beta\mu_C\,c\right].
\end{equation}
In this representation, a common tangent of $g$ selects the coexisting scaffold concentrations. A tangent touching three compositions would require at least two concave intervals between them. We looked for common tangents using the lower convex envelope and counted concave intervals on a finer grid in $\mu_C$. The scaffold grid extends into the dilute corner so that a phase with very low scaffold concentration can enter the comparison.

\begin{figure}[htbp]
\centering
\includegraphics[width=0.72\linewidth]{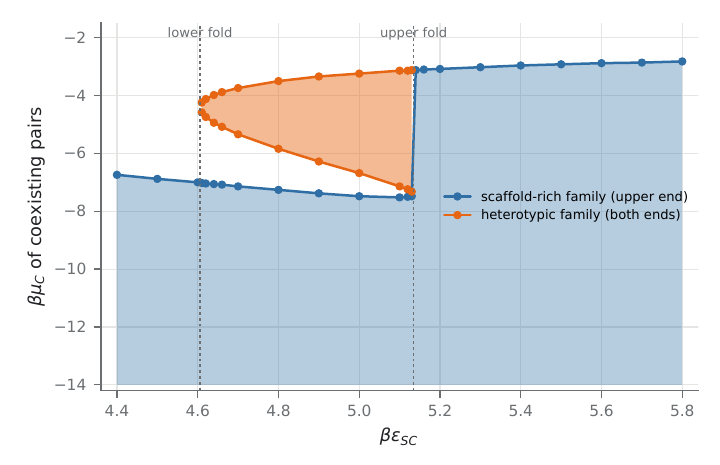}
\caption{Client chemical potentials of coexisting pairs for $A_6$--$C_2$ at $\beta\epsilon_{SS}=1.6$. The shaded bands show where each family has a pair. The scaffold-rich branch extends below the scan to the client-free limit. Between the folds (dashed lines), the heterotypic band lies at higher $\mu_C$ than the scaffold-rich band. Above the upper fold, the branches form one family.}
\label{fig:S-three-phase}
\end{figure}

At $\beta\epsilon_{SS}=1.6$, we scanned affinities on both sides of the folds. Each sampled point gave at most one coexistence pair and one concave interval on the respective grids. Between the folds, the scaffold-rich and heterotypic pairs occur in separate ranges of $\mu_C$ (Fig.~\ref{fig:S-three-phase}). At the upper fold these ranges meet and the two branches join. We also checked selected cuts at other scaffold cohesions. Wherever both families fell within the scan, their ranges of $\mu_C$ were separate. The scans found no three-phase coexistence on the cuts examined.

\FloatBarrier
\section{Critical points, fold continuation, and cusp scaling}
\label{si:critical-points}

The two folds in main-text Fig.~4 delimit the range of affinities with three interior critical points. We start from the conditions for a single critical point, then use them to locate the folds and their cusp. We use divalent clients and $v=1/8$ for both bond types unless stated otherwise.

On the incompressible lattice, specifying $\rho_S$ and $\rho_C$ fixes the solvent fraction. Two compositions coexist when they share a tangent plane to $f(\rho_S,\rho_C)$. Matching the slopes gives equal chemical potentials, and matching the intercepts gives equal osmotic pressure.
\begin{equation}
\nabla f(\boldsymbol\rho^{\rm I})=\nabla f(\boldsymbol\rho^{\rm II})\equiv\boldsymbol\mu,
\qquad
f(\boldsymbol\rho^{\rm I})-\boldsymbol\mu\cdot\boldsymbol\rho^{\rm I}
=f(\boldsymbol\rho^{\rm II})-\boldsymbol\mu\cdot\boldsymbol\rho^{\rm II}.
\label{eq:si-common-tangent}
\end{equation}
At a critical point, the two ends of a tie line approach the same composition. To find the direction in which they merge, write the endpoints as $\boldsymbol\rho^{\rm I,II}=\bar{\boldsymbol\rho}\mp r\boldsymbol u$ with $|\boldsymbol u|=1$. Let $H_{ij}=\partial^2 f/\partial\rho_i\partial\rho_j$ be the composition Hessian. Its eigenvalues $\lambda_-\le\lambda_+$ measure the curvature in the soft and stiff directions $\boldsymbol v$ and $\boldsymbol w$. We use subscripts to denote directional derivatives, for example $f_{vvv}=D_{\boldsymbol v}^3f$.

Subtracting the two chemical-potential equations and dividing by the endpoint separation $2r$ gives
\begin{equation}
H(\bar{\boldsymbol\rho})\,\boldsymbol u+\frac{r^2}{6}\,\partial^4 f[\boldsymbol u,\boldsymbol u,\boldsymbol u,\cdot]+O(r^4)=0,
\end{equation}
As the tie line shrinks, the leading term requires $H(\boldsymbol\rho_c)\boldsymbol u=\boldsymbol0$. The endpoints approach along the soft direction and $\lambda_-=0$, which places the critical point on the spinodal. The pressure equality gives $\bar{\boldsymbol\rho}\cdot H\boldsymbol u=0$ at the same order.

A stable critical point must also be a local minimum after subtracting its tangent plane. Write this difference as $\Omega(\boldsymbol\delta)=f(\boldsymbol\rho_c+\boldsymbol\delta)-f(\boldsymbol\rho_c)-\nabla f(\boldsymbol\rho_c)\cdot\boldsymbol\delta$. For $\boldsymbol\delta=t\boldsymbol v+s\boldsymbol w$, minimize over the stiff coordinate $s$ to let the composition relax. The resulting potential starts as $\widetilde\Omega(t)=\tfrac16 f_{vvv}t^3+\tfrac1{24}g_4t^4+O(t^5)$. The cubic term would lower the free energy on one side of the candidate critical point, so it must vanish.
\begin{equation}
\kappa_3\equiv f_{vvv}=D_{\boldsymbol v}^3f=0 .
\end{equation}
The quartic coefficient must be positive for an ordinary critical point. Its value after relaxation is
\begin{equation}
g_4=f_{vvvv}-\frac{3f_{vvw}^2}{\lambda_+}
\label{eq:si-g4}
\end{equation}
The second term accounts for the displacement $s=-f_{vvw}t^2/(2\lambda_+)$ along the stiff direction. The conditions $\lambda_-=0$ and $\kappa_3=0$ are equivalent to $\det H=0$ and $D_{\boldsymbol v}\det H=0$ when $\lambda_+>0$. We work directly with the soft eigenvalue to locate where this curvature vanishes.

\subsection{Formulation using the soft eigenvalue}
\label{si:lambda-form}
To follow critical points as affinity changes, it helps to view $\lambda_-$ as a function over the composition triangle. Its zero contour is the spinodal. Differentiating $\lambda_-=\boldsymbol v^{\mathsf T}H\boldsymbol v$ and using the normalization of $\boldsymbol v$ gives
\begin{equation}
\nabla\lambda_-=\kappa_3\,\boldsymbol v+f_{vvw}\,\boldsymbol w
\label{eq:si-gradlambda}
\end{equation}
A critical point is stationary along $\boldsymbol v$ on the zero contour. At a fold the curvature field is stationary in both composition directions. At the cusp this stationary point is degenerate.
\begin{center}
\begin{adjustbox}{max width=\linewidth}
\begin{tabular}{lll}
\toprule
object & condition & meaning for $\lambda_-(\boldsymbol\rho)$\\
\midrule
spinodal        & $\lambda_-=0$ & zero level set\\
critical point  & $\lambda_-=0$, $\;\nabla\lambda_-\cdot\boldsymbol v=0$ & level set stationary along the soft direction\\
fold            & $\lambda_-=0$, $\;\nabla\lambda_-=\boldsymbol0$ & a stationary point of $\lambda_-$ lies at level zero\\
cusp            & $\lambda_-=0$, $\;\nabla\lambda_-=\boldsymbol0$, $\;\det\nabla^2\lambda_-=0$ & that stationary point is degenerate\\
\bottomrule
\end{tabular}
\end{adjustbox}
\end{center}
This interpretation also follows from the equations we solve. For $\boldsymbol C=(\lambda_-,\kappa_3)$, the first row of the composition Jacobian $J=\partial\boldsymbol C/\partial\boldsymbol\rho$ is $f_{vvw}\boldsymbol w$ at a critical point. Thus $\det J=f_{vvw}(w_1\partial_2\kappa_3-w_2\partial_1\kappa_3)$, and at a generic fold $f_{vvw}=0$. We classify the stationary points by the signs of the eigenvalues of $\nabla^2\lambda_-$.
\begin{center}
\begin{adjustbox}{max width=\linewidth}
\begin{tabular}{llll}
\toprule
 & $\beta\epsilon_{SC}$ at $\beta\epsilon_{SS}=1.6$ & curvature signs & change in the spinodal\\
\midrule
lower fold & $\simeq 4.61$ & $(+,+)$ & a new loop appears\\
upper fold & $\simeq 5.13$ & $(-,+)$ & two loops meet\\
cusp       & --- & $(0,+)$ & the fold branches meet\\
\bottomrule
\end{tabular}
\end{adjustbox}
\end{center}
As affinity increases along the cut in main-text Fig.~4A, a minimum of $\lambda_-$ crosses zero first, followed by a saddle. This accounts for the appearance of a new spinodal loop and its later connection to the scaffold-rich loop. The two eigenvalues of $H$ remain distinct at these points and at the cusp, so $\lambda_-$ is smooth there.

\subsection{Locating critical points and folds}
\label{si:crit-numerics}
Every interior critical point lies on the spinodal. We trace $\lambda_-=0$ by interpolating the zero contour on a composition grid, then look along each branch for sign changes in $\kappa_3$. We orient $\boldsymbol v$ consistently along the branch because flipping it also flips the sign of $\kappa_3$. Each crossing gives a starting point for solving $(\lambda_-,\kappa_3)=\boldsymbol0$ by Newton iteration. We discard duplicate roots and treat the boundary separately. The count stabilizes on a $900\times900$ grid. Table~\ref{tab:S-cp} shows how it changes along the cuts in main-text Fig.~4A.

We locate a fold at fixed $\beta\epsilon_{SS}$ by adding $\det J=0$ to the critical-point equations and solving for the composition and $\beta\epsilon_{SC}$. To find the cusp, we also require $\boldsymbol\ell^{\mathsf T}\partial^2_{\boldsymbol\rho}\boldsymbol C[\boldsymbol u,\boldsymbol u]=0$, where $\boldsymbol u$ and $\boldsymbol\ell$ are right and left null vectors of $J$. When tracking the cusp with a model parameter, we solve for the null vector within the equations to keep its direction continuous along the branch.
\begin{equation}
\boldsymbol C=\boldsymbol0,\quad J\boldsymbol u=\boldsymbol0,\quad \boldsymbol u\cdot\boldsymbol u=1,\quad
J\boldsymbol y+\partial^2_{\boldsymbol\rho}\boldsymbol C[\boldsymbol u,\boldsymbol u]=\boldsymbol0,\quad \boldsymbol u\cdot\boldsymbol y=0,
\end{equation}
Here $\boldsymbol y$ allows the curvature along the null direction to be included in the continuation equations.

\begin{table}[htbp]
\centering\small
\caption{Interior critical points on the cuts of main-text Fig.~4A ($A_6$--$C_2$, $\beta\epsilon_{SS}=1.6$). $N_{\rm crit}$ is the number of interior
critical points, ``regions'' the number of disconnected coexistence regions.}
\label{tab:S-cp}
\begin{adjustbox}{max width=\linewidth}
\begin{tabular}{rccl}
\toprule
$\beta\epsilon_{SC}$ & $N_{\rm crit}$ & regions & critical compositions $(s,c)$\\
\midrule
4.4 & 1 & 1 & (0.1661, 0.0566) \\
4.7 & 3 & 2 & (0.0369, 0.1085); (0.0651, 0.0763); (0.1530, 0.0570) \\
5.125 & 3 & 2 & (0.0280, 0.1325); (0.1103, 0.0611); (0.1211, 0.0595) \\
5.3 & 1 & 1 & (0.0265, 0.1382) \\
\bottomrule
\end{tabular}
\end{adjustbox}
\end{table}

\subsection{Criticality on the boundaries of the composition triangle}
\label{si:edge-crit}
An interior critical point can leave the composition triangle as the interaction strengths change. We examine the edges to see where that can occur. On the scaffold-free edge, the association terms vanish and only ideal mixing remains. The soft curvature tends to $1/\rho_C+1/\rho_0>0$, which rules out a critical point on that edge. On the client-free edge, the free energy reduces to that of the binary scaffold--solvent mixture. The conditions $\lambda_-\to f''_{\rm bin}$ and $\kappa_3\to f'''_{\rm bin}$ place the edge critical point at $\beta\epsilon_{SS}\simeq1.37$ and $\rho_S\simeq0.29$. A scaffold-rich critical branch exists above this cohesion. On the solvent-free edge, the diverging curvature penalizes changes in total occupancy. With $H=\rho_0^{-1}\boldsymbol1\boldsymbol1^{\mathsf T}+K$ and $K$ finite, the soft mode follows the edge. Writing $\phi(x)=f(x,1-x)$ gives $\lambda_-\to\phi''(x)/2$. The edge critical points solve $\phi''=\phi'''=0$, tracing a curve from the binary scaffold critical point at zero client affinity. This boundary analysis explains how coexistence can persist with no interior critical point. At weak heterotypic affinity and strong scaffold cohesion, the coexistence band spans the client-free and solvent-free edges after its critical point leaves through the latter.

\subsection{Folds, cusp, and the \texorpdfstring{$3/2$}{3/2} law}
\label{si:cusp}

\subsubsection{Lyapunov--Schmidt reduction}
The cusp is where the two fold branches meet. To find how quickly they separate away from it, reduce the two criticality equations to a scalar equation along the direction in which the Jacobian is singular.

Write $\boldsymbol F(\boldsymbol x;\boldsymbol p)=(\lambda_-,\kappa_3)=\boldsymbol0$ for $\boldsymbol x=(\rho_S,\rho_C)$ and $\boldsymbol p=(\beta\epsilon_{SS},\beta\epsilon_{SC})$. A fold has singular $A=\boldsymbol F_{\boldsymbol x}$. At the cusp the quadratic term along its null direction also vanishes, $\boldsymbol\ell^{\mathsf T}\boldsymbol F_{\boldsymbol x\boldsymbol x}[\boldsymbol u,\boldsymbol u]=0$. Decompose a displacement from the cusp as $\delta\boldsymbol x=t\boldsymbol u+\boldsymbol\zeta$, with $\boldsymbol\zeta\perp\boldsymbol u$. We first solve for the perpendicular displacement from the range component of the equations. Its leading derivatives are $\boldsymbol\zeta_{p_i}=-A^{+}\boldsymbol F_{p_i}$ and $\boldsymbol\zeta_{tt}=-A^{+}\boldsymbol F_{\boldsymbol x\boldsymbol x}[\boldsymbol u,\boldsymbol u]$, where $A^+$ is the pseudoinverse.

Projecting the remaining equation onto $\boldsymbol\ell$ isolates the condition along $t$. With $\mu=\beta\epsilon_{SS}-\beta\epsilon^c_{SS}$ and $\nu=\beta\epsilon_{SC}-\beta\epsilon^c_{SC}$, it reads
\begin{equation}
g=\alpha_{SS}\mu+\alpha_{SC}\nu+\left(\beta_{SS}\mu+\beta_{SC}\nu\right)t+\frac{\kappa}{6}t^3+\dots,
\label{eq:si-reduced}
\end{equation}
\begin{equation}
\alpha_i=\boldsymbol\ell^{\mathsf T}\boldsymbol F_{p_i},\quad
\beta_i=\boldsymbol\ell^{\mathsf T}\!\left(\boldsymbol F_{\boldsymbol xp_i}\boldsymbol u+\boldsymbol F_{\boldsymbol x\boldsymbol x}[\boldsymbol u,\boldsymbol\zeta_{p_i}]\right),\quad
\kappa=\boldsymbol\ell^{\mathsf T}\!\left(\boldsymbol F_{\boldsymbol x\boldsymbol x\boldsymbol x}[\boldsymbol u,\boldsymbol u,\boldsymbol u]+3\,\boldsymbol F_{\boldsymbol x\boldsymbol x}[\boldsymbol u,\boldsymbol\zeta_{tt}]\right),
\end{equation}
The terms involving $\boldsymbol\zeta$ include the adjustment of composition perpendicular to $\boldsymbol u$. To measure the separation of the folds, subtract their common tangent of slope $m=-\alpha_{SS}/\alpha_{SC}$ by writing $\nu=m\mu+\eta$. Then $q=\beta_{SS}+m\beta_{SC}$ gives the coefficient of $\mu t$. At a fold, $g=\partial_tg=0$. The derivative condition gives $t^2=-2q\mu/\kappa$, and substituting this in $g=0$ gives $\eta_\pm=\pm(\kappa/3\alpha_{SC})(-2q\mu/\kappa)^{3/2}$. The two signs give the separation
\begin{equation}
\Delta\beta\epsilon_{SC}=P\left(\beta\epsilon^c_{SS}-\beta\epsilon_{SS}\right)^{3/2}+\dots,\qquad
P=\frac{2^{5/2}}{3\,|\alpha_{SC}|}\,\frac{|q|^{3/2}}{|\kappa|^{1/2}} .
\label{eq:si-32law}
\end{equation}
The $3/2$ power follows from the square-root displacement along the null direction and its cubic contribution to $g$. The prefactor follows from free-energy derivatives and is independent of the normalization of the null vectors.

\subsubsection{Numerical values and a parameter-free test}
For $A_6$--$C_2$ at $v=1/8$ the cusp is at
\begin{equation}
(\rho_S^c,\rho_C^c)\simeq(0.0657,\;0.0831),\qquad
(\beta\epsilon^c_{SS},\beta\epsilon^c_{SC})\simeq(1.737,\;4.526).
\end{equation}
The table gives the rounded coefficients from the reduction. The sign of $-2q/\kappa$ places the two folds on the side of weaker scaffold cohesion.
\begin{center}
\begin{adjustbox}{max width=\linewidth}
\begin{tabular}{lr lr}
\toprule
$\alpha_{SS}$ & $-1.639$ & $\beta_{SS}$ & $40.47$\\
$\alpha_{SC}$ & $-1.387$ & $\beta_{SC}$ & $-25.91$\\
$\kappa$ & $2.248\times10^4$ & $q$ & $71.08$\\
$m$ & $-1.181$ & $P$ & $5.4323$\\
\bottomrule
\end{tabular}
\end{adjustbox}
\end{center}
We check the prediction by tracing the two folds independently and dividing their measured separation by $\delta^{3/2}$, where $\delta=\beta\epsilon^c_{SS}-\beta\epsilon_{SS}$. The ratio $P_{\rm num}(\delta)$ should approach $P$ near the cusp. Combining results at two successive distances removes the leading error proportional to $\delta$ (Richardson extrapolation).
\begin{center}
\begin{adjustbox}{max width=\linewidth}
\begin{tabular}{rrr}
\toprule
$\delta$ & $P_{\rm num}(\delta)$ & Richardson\\
\midrule
$0.0005$ & $5.4413$ & \\
$0.0010$ & $5.4504$ & $5.4323$\\
$0.0020$ & $5.4686$ & $5.4322$\\
$0.0040$ & $5.5055$ & $5.4318$\\
\bottomrule
\end{tabular}
\end{adjustbox}
\end{center}
The extrapolation from the two points closest to the cusp agrees with the derivative-based prediction, $P\simeq5.4323$. The raw ratios drift as the distance increases, as expected from the higher-order terms.

\subsubsection{Quartic stability along the folds}
The fold describes a change in the number of critical points. To check that the critical points remain ordinary along the fold, we evaluate $\lambda_+$ and the relaxed quartic coefficient $g_4$. Both stay positive on the traced branches and at the cusp. On a fold, $f_{vvw}=0$ by Eq.~\eqref{eq:si-gradlambda}, so $g_4=f_{vvvv}$ there.

\begin{figure}[htbp]
  \centering
  \includegraphics[width=\linewidth]{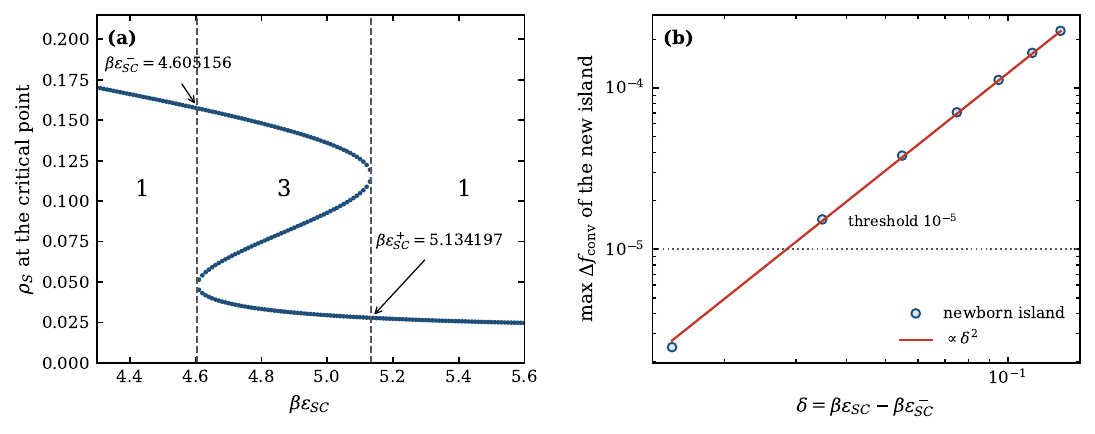}
  \caption{Critical-point folds and the onset of a new coexistence region for $A_6$--$C_2$ at $\beta\epsilon_{SS}=1.6$. (a) Dashed lines mark the two folds obtained from the critical-point equations. The number of interior critical points rises from one to three at the lower fold and returns to one at the upper fold. (b) Near the lower fold the maximum free-energy gap from the convex envelope grows as $\delta^2$, with $\delta=\beta\epsilon_{SC}-\beta\epsilon_{SC}^-$. A fixed gap threshold therefore registers the new region at a slightly higher affinity than the fold. The dotted line shows a threshold of $10^{-5}$.}
  \label{fig:S-folds}
\end{figure}

\FloatBarrier
\section{Robustness to molecular valence and scaffold cohesion}
\label{si:robustness}

We vary the bonding volume, site numbers, and scaffold cohesion to see which features of the phase diagram depend on the reference model. We track the fold conditions as the parameters change and count coexistence regions independently from the lower convex envelope. The latter step checks that the change in the critical-point count is accompanied by the expected change in coexistence.

\begin{figure}[htbp]
    \centering
    \includegraphics[width=\linewidth]{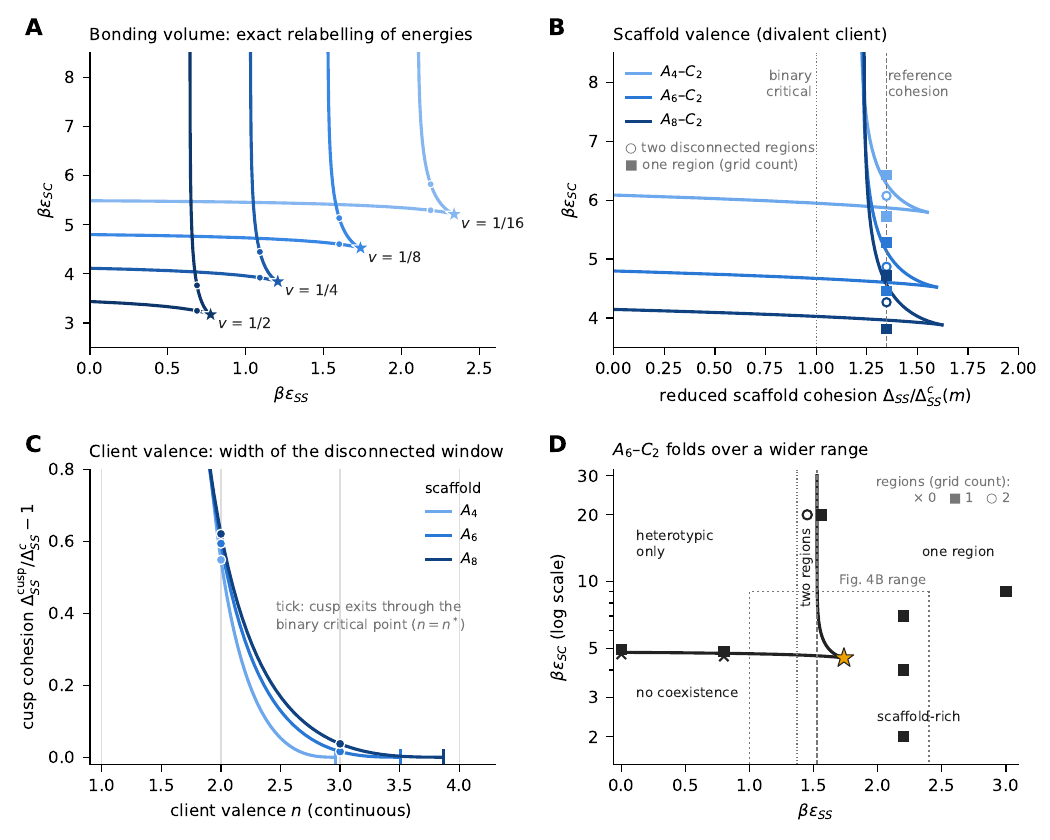}
    \caption{How the fold structure changes with bonding volume, valence, and cohesion. (A) Curves show the energy remapping of the $v=1/8$ folds for $A_6$--$C_2$. Stars and dots show the independently calculated cusps and folds. (B) Folds for divalent clients and different scaffold valences, with cohesion measured relative to the binary critical value. The numbered markers give the coexistence-region counts. (C) Circles mark integer client valences and ticks mark $n^*$, where the cusp leaves the composition interior. (D) Extended affinity range for $A_6$--$C_2$. The vertical lines mark the binary critical cohesion and the upper-fold asymptote. The box shows the range of main-text Fig.~4B.}
    \label{fig:S-robustness}
\end{figure}

Changing the bonding volume moves the energy coordinates according to Eq.~\eqref{eq:S-remap}. The independently calculated fold and cusp positions follow this mapping (Fig.~\ref{fig:S-robustness}A).

Scaffolds with different valences have different binary critical strengths. To compare their distances from scaffold-only phase separation, we measure cohesion relative to the binary critical value $\Delta_{SS}^{c}(m)$ for a scaffold with $m$ sites. At matched reduced cohesion, all three scaffold valences with divalent clients show one coexistence region before the lower fold, two between the folds, and one after reconnection (Fig.~\ref{fig:S-robustness}B).

To follow the cusp as client valence changes, we use $n$ as a continuous parameter between the integer valences. The cusp leaves the composition interior near $n=3$ for $A_4$, $3.5$ for $A_6$, and $3.9$ for $A_8$ (Fig.~\ref{fig:S-robustness}C). With trivalent clients, the window is absent for $A_4$ and narrow for $A_6$ and $A_8$. Refining the convex-envelope calculation confirms the two-region interval for $A_8$--$C_3$. The scans find no robust two-region interval for the monovalent or tetravalent clients over the tested ranges, $0.6\le\Delta_{SS}/\Delta_{SS}^{c}\le2.4$ and $0.5\le\beta\epsilon_{SC}\le12$. The low-valence end of the cusp is resolved between one and two client sites.

For $A_6$--$C_2$, we follow the branches from the cusp to find out whether the regions reconnect at stronger client affinity (Fig.~\ref{fig:S-robustness}D). Just above the binary scaffold critical cohesion, the regions remain separate throughout the tested range, which extends to $\beta\epsilon_{SC}=30$. At higher cohesion the upper fold occurs at finite affinity and the regions reconnect. Above the cusp the change is a smooth crossover. The convex-envelope counts agree with this sequence.

\FloatBarrier
\section{Monte Carlo model and analysis methods}
\label{si:mc}
\subsection{Model}
\label{si:mc-model}
The simulations retain the finite binding rules of the theory and place the molecules explicitly on a periodic body-centred cubic lattice. Each lattice site holds solvent, a scaffold, or a client. A scaffold has six arms and a $C_n$ client has $n$. Each arm points along one of the eight nearest-neighbour directions. A scaffold--scaffold or scaffold--client pair forms a bond when both molecules point an arm along the same edge. The energy is then $-\epsilon_{SS}$ or $-\epsilon_{SC}$. Each arm can bind one partner, and client--client contacts carry no attraction.

A species of valence $f$ has $g_f=\binom{8}{f}$ orientations. Giving each orientation weight $1/g_f$ normalizes the ideal contribution to the chemical potential across species. In the dilute limit, counting the aligned states of two molecules gives the association weight $W=(1/z_{\mathrm{nn}})\sum_{a,\gamma}n_an_\gamma[\exp(\beta\epsilon_{a\gamma})-1]$, where $z_{\mathrm{nn}}=8$ is the coordination number. Matching this two-particle weight to TPT1 gives $v=1/8$. The theory and simulations can then be compared at the same nominal energies.

\subsection{Sampling}
\label{si:mc-sampling}
To map a phase diagram, the simulation needs to visit many compositions. We use a semi-grand ensemble, in which a trial change of species can change the scaffold and client counts. With probability $p_{\rm swap}$, we exchange the identities and orientations of two sites to rearrange particles at the current composition. Otherwise a species change is proposed uniformly over solvent, scaffold, and client, and its orientation is drawn uniformly within that species. The Hastings factor cancels the normalization of the orientation weights. The acceptance probability includes the semi-grand fields and the umbrella bias. We record particle counts every four sweeps, with $V$ trial moves per sweep.

Without an umbrella, the system would spend most of its time at a few favoured compositions. We apply overlapping biases to sample the intervening compositions as well. Window $i$ adds
\begin{equation}
\beta u_i(\boldsymbol\phi)=\tfrac12\kappa_S(\phi_S-\phi^*_{S,i})^2+\tfrac12\kappa_C(\phi_C-\phi^*_{C,i})^2-V\left(\delta\mu_{S,i}\phi_S+\delta\mu_{C,i}\phi_C\right),
\end{equation}
The quadratic terms confine $\boldsymbol\phi=(N_S,N_C)/V$ near the target, with $\kappa_S=\kappa_C=17.5V$. The linear offsets $\delta\boldsymbol\mu_i$ move the mean to the target. We tune them in pilot simulations using $d\langle\boldsymbol\phi\rangle/d(\delta\boldsymbol\mu)=V\,{\rm Cov}(\boldsymbol\phi)$. We use the full covariance because changing either field can shift both concentrations. Targets are about two window widths apart, so nearby histograms overlap. Independent replicas start from random and compact configurations to sample from different initial arrangements.

We combine the histograms with the weighted histogram analysis method \cite{torrie1977,ferrenberg1989,kumar1992}. Including each window's full bias, including its linear field term, puts the windows on one free-energy surface $\beta F_V(N_S,N_C)$. The trivalent-client surfaces retain the integer particle counts. The divalent-client surfaces use 240 bins per axis and omit bins with fewer than 20 pooled observations.

\subsection{Finite-volume coexistence and tie lines}
\label{si:mc-coex}
In a finite box, a coexistence pair appears as two basins in the composition distribution. We adjust the semi-grand fields until the two basins have equal integrated probability \cite{borgs1990}. Equal weight makes neither basin preferred at those fields. The mean composition in each basin gives a tie-line endpoint.

For binned surfaces, the lower convex hull provides starting estimates of the fields. We keep well-separated endpoints (at least $0.04$ apart) with nearly equal weights and a projected barrier of at least $1\,kT$. Sparse sampling between the peaks can mimic a barrier. We therefore require at least one sampled state and 20 observations in every intervening bin, and at least 100 observations per basin. To estimate endpoint uncertainty we repeat the analysis while leaving out one replica at a time.

For trivalent clients we trace the equal-weight branch on the integer counts. We take smaller steps when the barrier falls below $3\,kT$ because the two basins become harder to distinguish as the tie line shrinks. We identify estimates that draw more than $1\%$ of their weight from within four composition-grid spacings of the support boundary or from states with fewer than 20 observations. The displayed branch includes these estimates where the projected modes remain stable.

Near the critical point, both sampling uncertainty and the choice of basin boundary can shift the endpoints. We resample complete replicas and compare with endpoints obtained by placing the basin boundary at the valley between the peaks. We require both the joint 95\% endpoint uncertainty radius and the shift from changing the basin boundary to be within $10^{-3}$ in composition. We omit isolated tie lines from the figure when their endpoints are more than $0.05$ from the main branch, both the edge and low-count probabilities exceed $0.99$, and fewer than ten effective observations remain after reweighting.

\FloatBarrier
\section{Monte Carlo tests of reentrance and phase organization}
\label{si:mc-phase-diagrams}
\subsection{Trivalent-client phase diagrams}
\label{si:mc-results}
\begin{figure}[htbp]
\centering
\includegraphics[width=\linewidth]{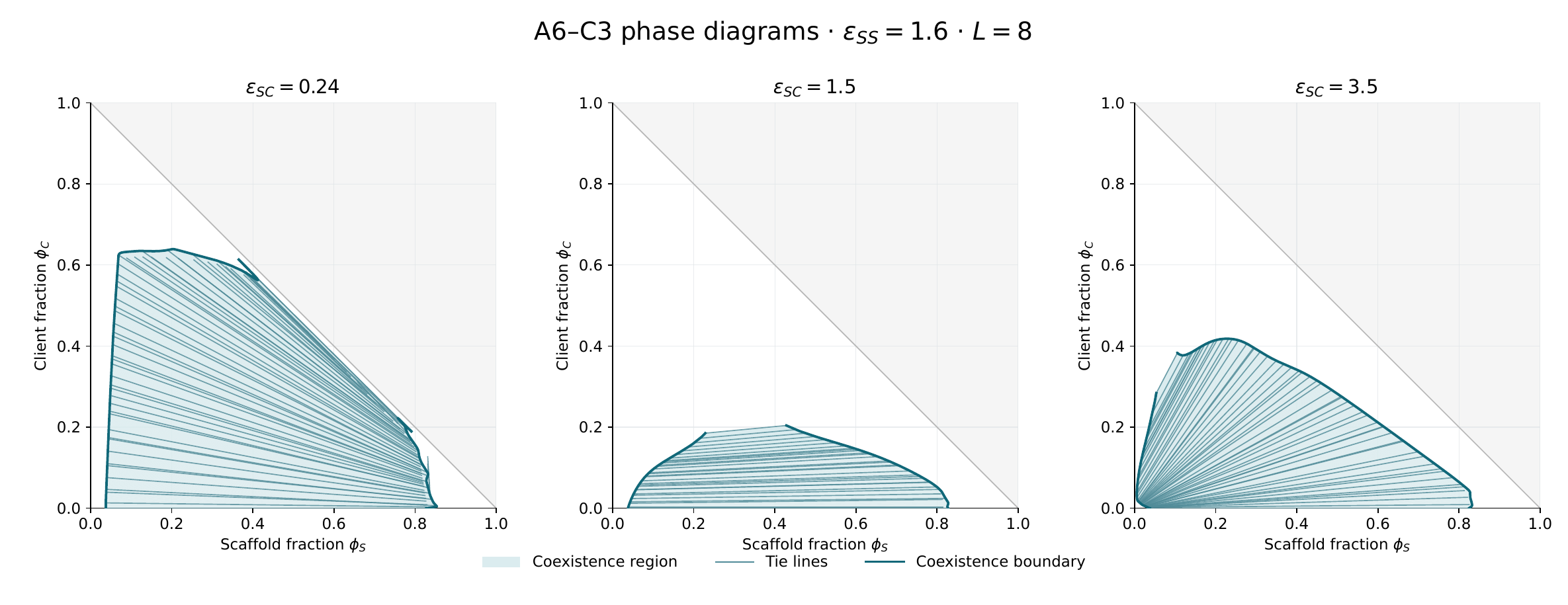}
\caption{Monte Carlo phase diagrams for $A_6$--$C_3$ at $\beta\epsilon_{SS}=1.6$ and $L=8$. From left to right, $\beta\epsilon_{SC}=0.24$, $1.5$, and $3.5$. Thin lines join the mean compositions of equal-weight basins. The shaded region follows the continued tie-line branch, bounded by the thick line. The diagonal marks $\phi_S+\phi_C=1$.}
\label{fig:S-mc-a6c3}
\end{figure}

We follow the parent $(0.20,0.20)$ through the three phase diagrams in Fig.~\ref{fig:S-mc-a6c3}. At weak affinity its tie line joins a scaffold-rich, client-poor phase to a client-rich, scaffold-poor phase. At intermediate affinity the boundary recedes to smaller client concentrations and the parent becomes homogeneous. At stronger affinity the parent separates again, now along an associative tie line. TPT1 gives the same sequence, with narrower coexistence regions at the same nominal energies.

\begin{figure}[htbp]
\centering
\includegraphics[width=\linewidth]{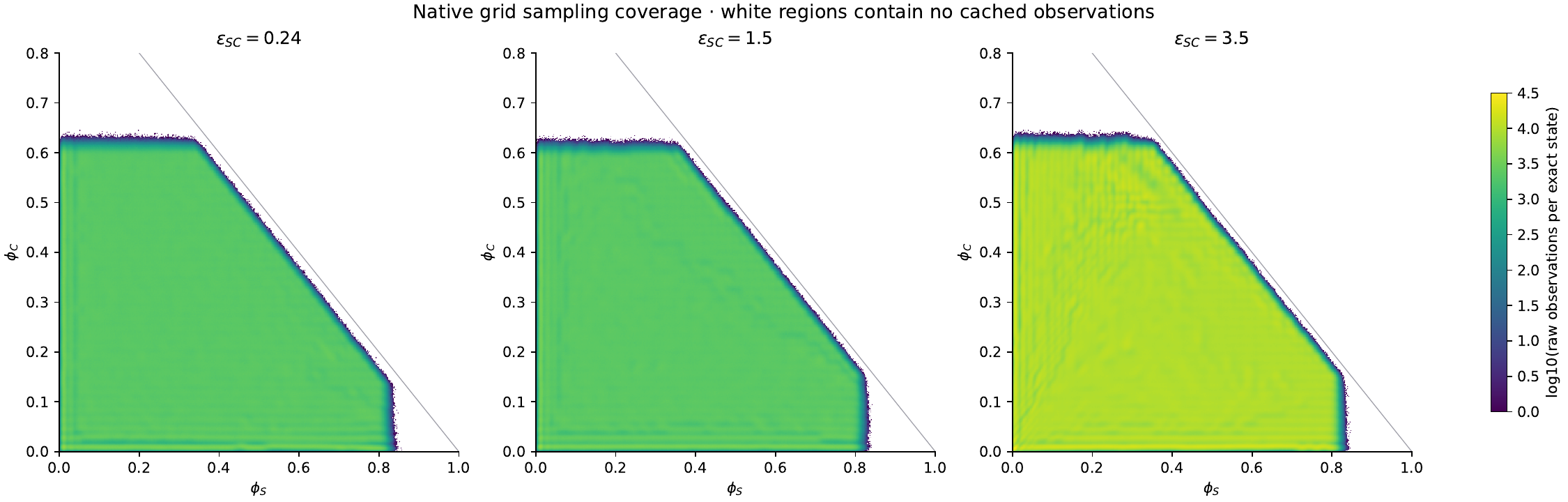}
\caption{Sampling coverage for the phase diagrams in Fig.~\ref{fig:S-mc-a6c3}. Colour shows the number of observations at each composition $(N_S,N_C)/V$ on a logarithmic scale, pooled over windows and replicas. White marks unsampled compositions.}
\label{fig:S-mc-coverage}
\end{figure}

The coverage map in Fig.~\ref{fig:S-mc-coverage} shows why we can trace the weak-affinity branch only partway. It reaches the upper client concentration sampled by the umbrellas ($\phi_C=0.6$), and the continuation depends on poorly sampled states. At the two higher affinities, the two basins become hard to distinguish as the tie lines shorten. Most of the near-critical endpoints fall short of the composition resolution specified in Sec.~\ref{si:mc-coex}.

\FloatBarrier

\subsection{Scaffold-affinity sweep for the divalent client}
\label{si:mc-sweep}
\begin{figure}[htbp]
\centering
\includegraphics[width=0.70\linewidth]{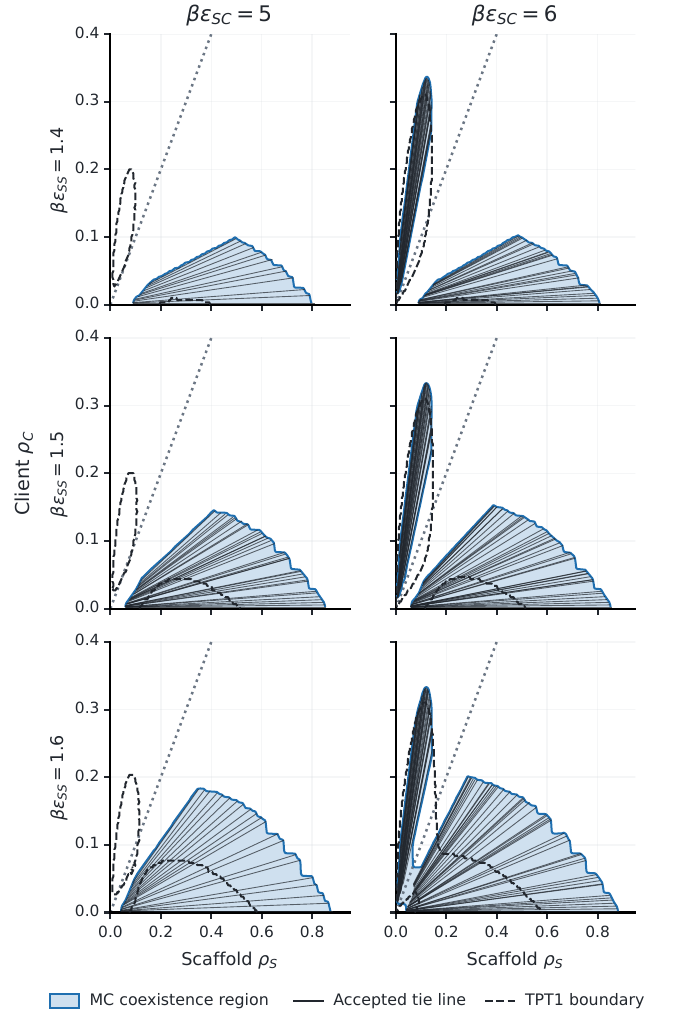}
\caption{Scaffold-affinity sweep for $A_6$--$C_2$ on the body-centred cubic lattice at $L=8$. Rows increase in scaffold cohesion and columns show $\beta\epsilon_{SC}=5$ and $6$. Shading follows the finite-volume tie lines (thin lines). The dashed curve shows the TPT1 coexistence boundary and the dotted line marks equal scaffold and client occupancies.}
\label{fig:S-mc-sweep}
\end{figure}

We vary scaffold cohesion to see whether the client-rich region joins the scaffold-rich region as readily when scaffold self-association is weaker. Figure~\ref{fig:S-mc-sweep} shows the two client affinities that bracket the change most clearly. The lower-affinity calculations show only the scaffold-rich family.

We use a common minimum barrier of $1\,kT$ for the binned surfaces so the panels use the same criterion for retaining tie lines. This includes pairs that satisfy the other selection criteria and meet this barrier threshold. To avoid drawing boundaries from poorly sampled endpoints, we omit pairs with both endpoints within $0.02$ of the client ceiling or both at $\phi_S+\phi_C>0.93$.

To draw the regions, we shade the cells swept by the retained tie lines, close gaps over a composition distance of about $0.0375$, and fill holes. We restrict the display to $\phi_S+\phi_C\le0.94$. We show components that contain a retained tie-line midpoint and have a free-energy gap above the convex hull of more than $4\,kT$. If none satisfy both criteria, we show the full swept region. The outline is the half-height contour after smoothing the indicator of this region.

At $\beta\epsilon_{SC}=5$, the scaffold-rich region extends towards larger client concentrations. This cut passes through the narrow TPT1 disconnection interval at $\beta\epsilon_{SS}=1.6$, where the finite-volume diagram still shows a single region. At $\beta\epsilon_{SC}=6$, a client-rich lobe with steep associative tie lines appears. It is separate from the scaffold-rich region at the two lower scaffold cohesions and connected at the highest cohesion shown. The simulated regions are broader than the TPT1 regions at the same nominal energies.

\FloatBarrier
\section{Monte Carlo tests of the affinity response}
\label{si:mc-response}
\subsection{Test of the bond-response identity}
\label{si:mc-identity}
The theory relates a free-energy derivative to the difference in bond counts between two states. In the finite box, we can test this relation by measuring a probability and a bond count independently. We use the parent near $(0.20,0.20)$ from main-text Fig.~2, with $N_S=N_C=102$ on $V=512$ sites. The reference $H=\{Q\le Q_0\}$ selects configurations with weak long-wavelength fluctuations while allowing local association. Here $Q$ is the sum of scaffold and client Fourier powers in the first two nonzero reciprocal-lattice shells, normalized by $V^2$. The free-energy cost per site of restricting the sample to $H$ is
\begin{equation}
G_{H,V}=-\frac{1}{V}\ln P_{\rm eq}(H)=\frac{1}{V}\left(\ln Z-\ln Z_H\right),
\end{equation}
The probability is $Z_H/Z$, the ratio of the restricted and unrestricted partition functions. To take the derivative, count aligned contacts of type $ij$ as $M_{ij}$, so $\beta U=-\beta\epsilon_{SS}M_{SS}-\beta\epsilon_{SC}M_{SC}$. With $\lambda=\ln[\exp(\beta\epsilon_{SC})-1]$, differentiating the two log partition functions gives
\begin{equation}
\frac{dG_{H,V}}{d\lambda}=\frac{\left(1-e^{-\beta\epsilon_{SC}}\right)\left(\langle M_{SC}\rangle_{\rm eq}-\langle M_{SC}\rangle_H\right)}{V},
\label{eq:S-mc-identity}
\end{equation}
The two averages compare the whole equilibrium ensemble with the subset $H$. The factor $1-e^{-\beta\epsilon_{SC}}$ converts the contact-count derivative to a derivative in $\lambda$, the affinity variable used for the association term. We use the same cutoff $Q_0=0.015$ at every affinity and for both architectures. The two nearby cutoffs, $0.012$ and $0.018$, test how sensitive the result is to the choice of reference. The independent-pool lattice model assigns separate six-arm sets to $S$--$S$ and $S$--$C$ binding on the same eight-neighbour lattice.

The reference can be rare under equilibrium sampling. We use twelve overlapping harmonic umbrellas in $Q$ to measure both the reference and the rest of the distribution. Their centres are spaced by $0.006$, starting at $-0.012$, and their stiffness is $4\times10^4$. Particle exchanges and arm reorientations have probabilities $0.6$ and $0.4$, respectively. Replica exchanges between umbrellas improve sampling across $Q$. After equilibration, multistate Bennett acceptance ratio (MBAR) reweighting removes the umbrella biases. We apply the reference cutoff to the reconstructed distribution.

To test the derivative, we estimate $G_{H,V}$ at the midpoint $\lambda_0$ and at $\lambda_0\pm h$ for two step sizes. We use $h=0.1$ and $0.05$ to check how the numerical slope depends on the step. A separate set of simulations at the midpoint measures the bond contrast, so the slope and contrast can be compared with independent uncertainties. We use four ladders per affinity and measurement, split between random and compact starts.

We check that the umbrellas form a connected set with histogram overlap of at least $0.03$ on each connecting edge. This allows their relative weights to be estimated. We require at least 50 effective observations per window and 100 in each target ensemble, including $H$, so each average has adequate sampling. We also compare halves of the trajectories and different starting configurations to detect incomplete equilibration. For the uncertainty, we resample ladders and circular blocks within each ladder so that the estimates account for both independent runs and time correlations. The sampling checks pass for all conditions.

\begin{figure}[htbp]
\centering
\includegraphics[width=0.92\linewidth]{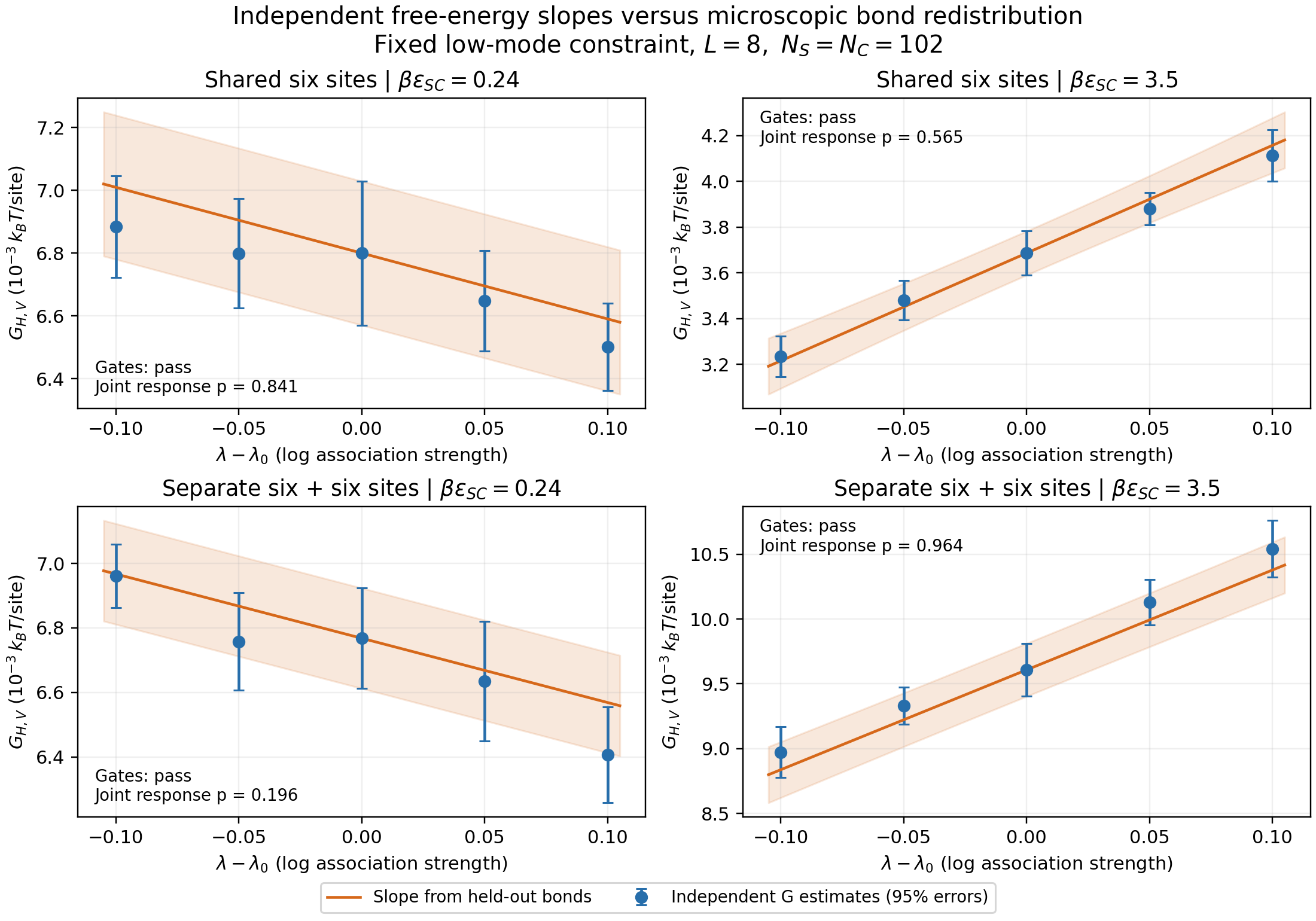}
\caption{Independent measurements of the probability gap $G_{H,V}$ and its slope. Columns show weak and strong client affinity. Rows show shared and separate site pools. Points show the measured gaps with bootstrap error bars. The line is drawn through the independently measured midpoint gap with slope set by the bond contrast. The band includes uncertainty in both.}
\label{fig:S-mc-identity}
\end{figure}

Table~\ref{tab:S-mc-identity} and Fig.~\ref{fig:S-mc-identity} show the comparison. Both architectures give a negative response at weak affinity and a positive response at strong affinity, and the slope estimates agree with the independently measured bond contrasts within uncertainty. Moving the reference cutoff changes the size of $G_{H,V}$ while preserving the signs and the agreement.

\begin{table}[htbp]
\centering\small
\caption{The measured free-energy slope and the independent bond contrast for the two binding architectures. The parent has $N_S=N_C=102$ and $L=8$, with $\beta\epsilon_{SS}=1.6$ and $Q_0=0.015$. Free energies, slopes, and contrasts are in units of $10^{-3}$ per site. The joint $p$ value compares both finite-difference slopes with the bond contrast.}
\label{tab:S-mc-identity}
\begin{adjustbox}{max width=\linewidth}
\begin{tabular}{lrrrrrr}
\toprule
architecture & $\beta\epsilon_{SC}$ & $G_{H,V}$ & $dG/d\lambda$, $h=0.1$ & $dG/d\lambda$, $h=0.05$ & bond contrast & joint $p$\\
\midrule
shared sites & 0.24 & $6.800\pm0.117$ & $-1.912\pm0.544$ & $-1.508\pm1.207$ & $-2.095\pm0.059$ & 0.84\\
shared sites & 3.5 & $3.687\pm0.049$ & $4.397\pm0.367$ & $3.999\pm0.569$ & $4.710\pm0.369$ & 0.56\\
separate sites & 0.24 & $6.768\pm0.079$ & $-2.769\pm0.455$ & $-1.232\pm1.218$ & $-1.993\pm0.062$ & 0.20\\
separate sites & 3.5 & $9.607\pm0.104$ & $7.840\pm0.749$ & $7.984\pm1.151$ & $7.704\pm0.346$ & 0.96\\
\bottomrule
\end{tabular}
\end{adjustbox}
\end{table}

\FloatBarrier

\subsection{Trivalent-client affinity scan and size extrapolation}
\label{si:mc-trivalent}
Figure~3F asks whether clients can enrich in the dense region while the separated sample has fewer heterotypic bonds than its reference. We use the shared-site $A_6$--$C_3$ model at $\beta\epsilon_{SS}=1.6$. To see how the comparison changes with size, we use $L=8$, $10$, $12$, and $16$ with $V=L^3$. Particle counts are rounded to give the same nominal parent, $(s_0,c_0)\simeq(0.33,0.0195)$. The plotted affinities run from $u_{SC}=\beta\epsilon_{SC}=0$ to $2$.

We sample each ensemble directly at fixed particle counts, using four independent chains. Particle exchanges rearrange the mixture and arm reorientations change the bonding pattern. We use the same move probabilities as in the identity test. Moves use Metropolis acceptance. In the reference runs we also reject moves that leave $H$. We discard the first 5\,000 sweeps for equilibration and measure the bond counts and the scaffold density pattern in the configurations that follow.

We distinguish separated states from the reference by looking at the scaffold distribution alone. This leaves client partitioning free to respond to the scaffold pattern. We measure long-wavelength scaffold fluctuations using
\begin{equation}
Q_S=\frac{1}{V^2}\sum_{\mathbf k\in\mathcal M}
\left|\sum_{i\in S}e^{i\mathbf k\cdot\mathbf r_i}\right|^2,
\label{eq:S-mc-c3-Q}
\end{equation}
Here $\mathcal M$ contains one representative from each opposite-wavevector pair in the first two nonzero reciprocal shells. The reference cutoff $Q_S\le0.012(8/L)^3$ keeps $VQ_S$ bounded by the same value at every size. This accounts for the $1/V$ scale of the Fourier power in a homogeneous phase with short-range correlations. For the separated state, we average over configurations from the unrestricted runs with $Q_S\ge0.025$. We vary this threshold by 10\% in either direction to check the sensitivity to the state definition.

For each separated configuration, the strongest scaffold Fourier mode locates the dense and dilute regions even if they move within the box. Its phase defines the coordinate $\theta$. Sites with $\cos\theta>0.5$ are assigned to the dense core and those with $\cos\theta<-0.5$ to the dilute core, leaving the middle region out of the concentration comparison. For each core we divide the mean client count by the mean core volume. For the bond contrast we count contacts throughout the box, since bonds in the middle region also contribute to the free-energy response.
\begin{align}
K_C&=\frac{\langle N_{C,D}\rangle_{\rm Sep}/\langle V_D\rangle_{\rm Sep}}
{\langle N_{C,L}\rangle_{\rm Sep}/\langle V_L\rangle_{\rm Sep}},\notag\\
\Delta m_{SC}&=\frac{\langle M_{SC}\rangle_{\rm Sep}-\langle M_{SC}\rangle_H}{V},
\qquad \Delta b_{SC}^{\rm MC}=(1-e^{-u_{SC}})\Delta m_{SC}.
\label{eq:S-mc-c3-measurements}
\end{align}
At zero affinity the Mayer factor makes the bond contrast vanish. Site exclusion still affects where the clients sit. The measured core ratio at zero affinity agrees with the ratio of the fractions of sites available to clients in the two cores.

We use direct estimates from chains that pass the sampling checks. At the plotted affinity $u_{SC}=1.75$, the $L=10$ and $12$ estimates come from reweighting simulations within $0.05$ on either side of that affinity. The factor $\exp[(u_{SC}-u_{\rm anchor})M_{SC}]$ changes the statistical weight of each configuration to match the target affinity. We apply it separately in $H$ and in the separated ensemble. We require at least ten contributing configurations and an effective sample size of at least 80 and at least a quarter of its unweighted value. The largest normalized weight can be at most $0.03$, so a single configuration cannot dominate. We average the two estimates with weights proportional to their effective sample sizes once they agree within $3.5$ standard errors.

The points in Fig.~3F are the intercepts of weighted fits in $1/L$. We give more weight to the more precise estimates and resample chains and blocks of time points through the full calculation, including reweighting and fitting. This propagates uncertainty from the simulations to the pointwise 95\% intervals on the intercepts. At $u_{SC}=1.5$, the fitted core ratio is about $1.34$ and the bond contrast is negative. The interval includes zero near the sign change and the contrast is positive at the strongest affinity shown. At $1.5$, the interval remains negative if we omit the smallest box or use a $1/L^2$ correction alone. It spans zero if we fit both correction terms. The comparison throughout uses the specified sequence of constrained references.

Figure~3F uses the core ratio $K_C^{\rm core}$. In a given box it equals the bulk ratio $K_C^{\rm bulk}=c_D/c_L$ when both cores lie on the bulk concentration plateaus. To see how an interface can change the measured ratio, let $\eta_L$ and $\eta_D$ be the effective admixtures of opposite-phase material in the two cores. For a profile that stays between the plateau values, write $\Delta c=c_D-c_L$ to obtain
\begin{align*}
c_L^{\rm core}&=c_L+\eta_L\Delta c,\qquad c_D^{\rm core}=c_D-\eta_D\Delta c,\\
K_C^{\rm core}&=K_C^{\rm bulk}\,
\frac{1-(1-1/K_C^{\rm bulk})\eta_D}{1+(K_C^{\rm bulk}-1)\eta_L}
\le K_C^{\rm bulk}\qquad (K_C^{\rm bulk}>1).
\end{align*}
For comparable admixtures, the dilute core has the larger relative bias because it starts with fewer clients per site. The inequality assumes concentrations stay between the two bulk values.

For a periodic slab, each core spans one third of the repeat distance $\ell$ along the normal. If $\phi$ is the dense-phase fraction, the margins from the core edges to the nominal interfaces are $d_D=\ell(\phi/2-1/6)$ and $d_L=\ell[(1-\phi)/2-1/6]$. A core samples a plateau to the desired accuracy once its margin exceeds the interfacial width $w$. At the representative affinity, the TPT1 dense fraction is about $0.54$. This requires a period of roughly ten interfacial widths for the dense core and sixteen for the dilute core. The core bias depends on the concentration profile. Beyond a tanh-like interface its tail is of order $\exp(-2d/w)$, so the linear $1/L$ fit for the core ratio is empirical. The bond count includes the entire box. In a slab of fixed shape, a size-independent excess of bonds per unit interfacial area gives a contribution proportional to area over volume, or $1/L$. The total contrast also includes the reference, which varies with size.

The tie-line slope gives a second way to compare the simulation with theory. If the scaffold and client profiles move along the same tie line through the interface, mixing reduces both core concentration differences by the same factor. Their ratio $t_{\rm core}=(c_D^{\rm core}-c_L^{\rm core})/(s_D^{\rm core}-s_L^{\rm core})$ then equals the bulk slope. At $u_{SC}=1.5$, the largest box gives a core enrichment of about $1.28$, compared with the TPT1 bulk value of $1.57$. Across sizes, the core slopes are near $0.01$, about half the TPT1 slope. Mixing along the TPT1 tie line would leave its slope unchanged. In the largest box, the material between the cores also contains slightly more client than predicted by the line through the two core compositions. The aligned concentration profiles can help distinguish interfacial enrichment from a shift in the bulk endpoints.

\FloatBarrier

\end{document}